\documentclass{amsart}
\usepackage{amssymb,latexsym}
\usepackage{epsfig}
\usepackage{eufrak}
\usepackage{amsmath}
\usepackage{mathrsfs}
\usepackage{color}

\theoremstyle{plain}
\newtheorem{theorem}{Theorem}

\newtheorem{lemma}{Lemma}

\theoremstyle{definition}
\newtheorem{definition}{Definition}

\theoremstyle{remark}

\numberwithin{equation}{section}

\newcommand{\be}{\begin{equation}}
\newcommand{\ee}{\end{equation}}
\newcommand{\bee}{\begin{eqnarray*}}
\newcommand{\eee}{\end{eqnarray*}}
\newcommand{\bel}{\begin{eqnarray}}
\newcommand{\eel}{\end{eqnarray}}
\newcommand{\bec}{\begin{cases}}
\newcommand{\eec}{\end{cases}}
\newcommand{\bem}{\begin{bmatrix}}
\newcommand{\eem}{\end{bmatrix}}

\newcommand{\la}{\label}
\newcommand{\li}{\left}
\newcommand{\ri}{\right}

\newcommand{\DEF}{\stackrel{\mathrm{def}}{=}}

\newcommand{\lc}{\lceil}
\newcommand{\rc}{\rceil}
\newcommand{\lf}{\lfloor}
\newcommand{\rf}{\rfloor}

\newcommand{\ep}{\epsilon}
\newcommand{\vep}{\varepsilon}
\newcommand{\lm}{\lambda}

\newcommand{\Up}{\Upsilon}

\newcommand{\si}{\sigma}

\newcommand{\de}{\delta}
\newcommand{\De}{\Delta}
\newcommand{\vDe}{\varDelta}

\newcommand{\ga}{\gamma}

\newcommand{\se}{\theta}

\newcommand{\al}{\alpha}
\newcommand{\ba}{\beta}

\newcommand{\ro}{\rho}

\newcommand{\f}{\frac}
\newcommand{\sq}{\sqrt}
\newcommand{\cd}{\cdots}
\newcommand{\qu}{\quad}
\newcommand{\qqu}{\qquad}
\newcommand{\fa}{\forall}

\newcommand{\mscr}{\mathscr}
\newcommand{\mcal}{\mathcal}

\newcommand{\bb}{\mathbb}

\newcommand{\wh}{\widehat}

\newcommand{\mrm}{\mathrm}
\newcommand{\bs}{\boldsymbol}

\newcommand{\ap}{\approx}

\newcommand{\sh}{\slash}

\newcommand{\tx}{\text}

\newcommand{\iy}{\infty}

\newcommand{\pa}{\partial}

\newcommand{\bed}{\begin{description}}
\newcommand{\eed}{\end{description}}
\newcommand{\bei}{\begin{itemize}}
\newcommand{\eei}{\end{itemize}}
\newcommand{\ben}{\begin{enumerate}}
\newcommand{\een}{\end{enumerate}}
\newcommand{\bib}{\bibitem}
\newcommand{\beL}{\begin{lemma}}
\newcommand{\eeL}{\end{lemma}}
\newcommand{\beT}{\begin{theorem}}
\newcommand{\eeT}{\end{theorem}}
\newcommand{\sect}{\section}

\newcommand{\bpf}{\begin{pf}}
\newcommand{\epf}{\end{pf}}
\newcommand{\bsk}{\bigskip}

\newcommand{\pfbox}{\hfill\mbox{$\Box$}}
\newenvironment{pf}{\paragraph*{Proof{\rm.}}}{\pfbox\bigskip}

\begin{document}

\title[A Statistical Theory for the Analysis of Uncertain Systems] {A Statistical Theory for the Analysis of Uncertain Systems}

\author{Xinjia Chen, Kemin Zhou and Jorge L. Aravena}

\address{Department of Electrical and Computer Engineering\\
Louisiana State University\\
Baton Rouge, LA 70803}

\email{chan@ece.lsu.edu\\
kemin@ece.lsu.edu\\
aravena@ece.lsu.edu}

\thanks{This
research was supported in part by grants from NASA (NCC5-573), LEQSF (NASA /LEQSF(2001-04)-01), the NNSFC Young
Investigator Award for Overseas Collaborative Research (60328304) and a NNSFC grant (10377004).}


\date{March 2007}

\begin{abstract}

This paper addresses the issues of conservativeness and
computational complexity of probabilistic robustness analysis. We
solve both issues by defining a new sampling strategy and
robustness measure. The new measure is shown to be much less
conservative than the existing one. The new sampling strategy
enables the definition of efficient hierarchical sample reuse
algorithms that reduce significantly the computational complexity
and make it independent of the dimension of the uncertainty space.
Moreover, we show that there exists a one to one correspondence
between the new and the existing robustness measures and provide a
computationally simple algorithm to derive one from the other.

\end{abstract}

\maketitle

\sect{Introduction}

{\em Robustness analysis} is used to predict if a system will
perform satisfactorily in the presence of uncertainties. It is
generally accepted as an essential step in the design of
high-performance control systems. In practice, the analysis has to
be very efficient because it has to use models as realistic as
possible and, usually, it takes many cycles of analysis-design to
come up with a satisfactory controller. The outcome of the
robustness analysis should allow the designer not only to evaluate
the robust performance of a controller, but also to compare
various controllers in order to obtain the best control strategy.
Needless to say, unnecessary conservativeness prevents a realistic
analysis.

Aimed at overcoming the computational complexity and conservatism
of the classical deterministic worst-cast approach, there are
growing interests in developing probabilistic methods and
randomized algorithms (see, \cite{bai}-\cite{C2},
\cite{Kan}-\cite{Wang} and the references therein). Specially, a
probabilistic robustness measure, referred to as the {\it
confidence degradation function} or {\it robustness function} is
proposed in \cite{BLT}. Such robustness measure has been
demonstrated to be much superior  than the classical deterministic
robustness margin in terms of conservatism, computational
complexity and generality of application.

The computation of the {\em robustness function} using Monte Carlo
simulations requires {\it uniform sampling} from bounding sets in
the uncertainty space, which can reach high dimensions very
quickly; for example if the uncertainty is modelled by a $5\times
5$ complex-valued matrix then the dimension of the uncertainty
space is 50. We will show here that such sampling suffers from
what we term {\em surface effect} and may introduce undue
conservativeness in the evaluation of system robustness. We
address this conservativeness with a new sampling technique and a
new probabilistic robustness measure that is significantly less
conservative. Moreover, with a suitable computing structure it can
be evaluated for arbitrarily dense gridding of uncertainty radius
with a computational complexity that is very low and {\em is
independent of the dimension of uncertainty.}

We shall use the following notation throughout this paper. The
uncertainty is denoted as boldface $\bs{\vDe}$ and its realization
is denoted as $\vDe$.  The probability density function of
$\bs{\vDe}$ is denoted as $f_{\bs{\vDe}}$. We measure the size of
uncertainty by a function $||.||$ which has the {\it scalable
property} that $||\ro \vDe|| = \ro || \vDe||$ for any uncertainty
instance $\vDe$ and any $\ro > 0$.  Obviously, the most frequently
used $H_\iy$ or $l_p$ norm of uncertainty possesses such scalable
property. The uncertainty bounding set of radius $r$ is denoted as
$\mcal{B}_r = \{ \vDe: ||\vDe|| \leq r \}$.  We use $\pa \mcal{B}_r$
to denote $\{ \vDe: ||\vDe|| = r \}$.  Specially, $\mcal{B}$ denotes
$\{ \vDe: ||\vDe|| \leq 1 \}$ and $\pa \mcal{B}$ denotes $\{ \vDe:
||\vDe|| = 1 \}$. For a subset $S_r$ of $\pa \mcal{B}_r$, its
``area'' is defined as \be \la{defarea}
 \mrm{area}(S_r) = \lim_{\vep_1
\downarrow 0 \atop{ \vep_2 \downarrow 0 } } \f{ \int_{q \in \li
\{\f{\ro}{r} \vDe : \; r - \vep_1 \leq \ro \leq r + \vep_2 , \; \vDe
\in S_r \ri \} } d q } { \vep_1 + \vep_2 } \ee where ``$\int$''
denotes the multivariate Lebesgue integration and the down arrow
``$\downarrow$'' means ``decreases to''.

The indicator function $\bb{I} (.)$ means that $\bb{I} (\vDe) = 1$
if the robustness requirement is guaranteed for $\vDe$ and $\bb{I}
(\vDe) = 0$ otherwise. The probability of an event is denoted as
$\Pr \{. \}$.   The conditional probability is denoted as $\Pr \{.
\mid . \}$.  The set of complex number is denoted as $\bb{C}$. The
set of real matrices of size $m \times p$ is denoted as $\bb{R}^{m
\times p}$. The set of complex matrices of size $m \times p$ is
denoted as $\bb{C}^{m \times p}$.  The real and complex parts of a
number is denoted as $\Re(.)$ and $\Im(.)$ respectively. The largest
and the second largest singular values of a matrix are denoted as
$\overline{\si}(.)$ and $\si_2(.)$ respectively.  The ceiling
function is denoted as $\lc . \rc$ and the floor function is denoted
as $\lf . \rf$.

\subsection{The Surface Effect of Uniform Sampling}
In order to illustrate the {\em surface effect}, consider a {\it
uniform sampling} extracting samples from the uncertainty set
$\mcal{B}_r$. Let $E_\rho$ denote the event that {\it a sample
chosen uniformly from $\mcal{B}_r$ lies outside the bounding set
$\mcal{B}_\rho$ of radius $\ro < r$}. Under the assumption of
uniform distribution it is easy to see that such event will have the
probability $\Pr \{E_\rho\} = 1-\left(\frac{\rho}{r}\right )^d$
where $d$ is the dimension of uncertainty.  As $d$ increases this
probability approaches one for all $\rho<r$. For example when
$\f{\rho}{r}=0.9$ and $d=50$ then $\Pr \{E_\rho \}=0.9948$. {\bf
Hence out of 1000 samples extracted uniformly from the bounding set
of radius $r$ one would expect that about 995 will be outside the
bounding set with radius $\rho = 0.9 r$}. If the uncertainty is well
modeled one can reasonably assume that large uncertainties are less
likely than small ones and we are faced with the fact that the
uniform sampling selects cases that are not indicative of the actual
situation but present a very unfavorable picture. In Section 2 we
discuss in detail the modeling of uncertainties and show that
uniform sampling can give a very conservative evaluation of system
robustness. In Section 3 we introduce a new sampling technique and a
new robustness measure which overcomes the conservativeness issue.
Section 4 establishes a one to one mapping between our measure and
the existing one and considers other capabilities of the new
robustness function.  The detail algorithms are presented in Section
5.  Section 6 addresses the issue of computational complexity for
the evaluation of robustness function. In particular we show that by
using a special type of {\em hierarchial} data structure it is
possible to design computational algorithms that have a complexity
that is independent of the dimension of the uncertainty.  The proofs
of theorems are given in the Appendix.

\section{Modeling Uncertainty}

In this section, we shall discuss the characteristics of uncertainty from the perspective of modelling
practices.
\begin{figure}
\centering
\includegraphics[height=3cm]{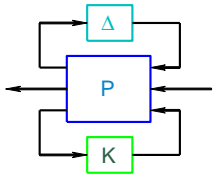}
\caption{Uncertain System} \label{fig_Diag_Sys}
\end{figure}

Consider an uncertain system shown in Figure \ref{fig_Diag_Sys}. In
control engineering, one usually takes into account all possible
directional information about the uncertainty by introducing
weighting matrices and absorbing it into the generalized plant $P$.
Therefore, it is reasonable to assume that the uncertainty
$\bs{\vDe}$ is {\it radially symmetrical} in distribution in the
sense that, for any $r > 0$ and any $S_r \subseteq \{ \vDe :
||\vDe|| = r  \}$,
\[
\Pr \{  \bs{\vDe} \in S_r \mid ||\bs{\vDe}|| = r \} = \f{
\mrm{area}(S_r) } {\mrm{area} (\pa \mcal{B}_r) }
\]
if $f_{||\bs{\De}||}(.)$ is continuous at $r$, where the conditional
probability in the left hand side is defined as \[ \lim_{\vep_1
\downarrow 0 \atop{ \vep_2 \downarrow 0 } } \f{ \Pr \li \{ \bs{\vDe}
\in \{ \vDe \in S_\ro: \; r - \vep_1 \leq \ro \leq r + \vep_2  \}
\ri \} } { \Pr \{ r - \vep_1 \leq ||\bs{\vDe}|| \leq r + \vep_2 \}
}.
\]
On the other hand, one usually attempts to make the magnitude of
modelling error, measured by $||\bs{\vDe}||$, as small as possible.
Due to the effort to minimize $||\bs{\vDe}||$ in modelling, it is
reasonable to assume that {\it small modelling error is more likely
than large modelling error}.  This gives rise to the rationale of
treating $||\bs{\De}||$ as a random variable such that its density,
$f_{||\bs{\De}||}(r) = \f{d [\Pr \{ ||\bs{\De}|| \leq r \}] } {d r
}$, is {\it non-increasing} with respect to $r$.  In the sequel, we
shall use $\mscr{F}$ to denote the family of radially symmetrical
and non-increasing density function $f_{\bs{\vDe}}$. It should be
noted that a wider class of probability density functions, denoted
by $\mscr{G}$, has been proposed in \cite{BLT} to model uncertainty.
Such family $\mscr{G}$ consists of radially symmetrical density
function $f_{\bs{\vDe}}$ that is non-increasing in the sense that
$f_{\bs{\vDe}} (\vDe_1) \leq f_{\bs{\vDe}} (\vDe_2)$ if $||\vDe_1||
\geq ||\vDe_2||$. It can be shown that $\mscr{G}$ is a superset of
$\mscr{F}$, i.e., $\mscr{G} \supseteq \mscr{F}$ (see Lemma 2 in
Appendix A).

\subsection{Existing Robustness Function} \la{3A}

The existing robustness function, proposed in \cite{BLT}, is given by
\[
\underline{\bb{P}} (r) \DEF \inf_{\ro \in (0, r]} \bb{P} (\ro) \]
 with
 \[
 \bb{P} (r) = \Pr \{ \bb{I}(\bs{\vDe}^\mrm{u} )
 = 1 \}
 \]
  where $\bs{\vDe}^\mrm{u}$ is uniformly distributed over $\mcal{B}_r$.
  It has been shown in \cite{BLT} that $\underline{\bb{P}} (r)$
 is a lower bound of the probability of guaranteeing the robustness requirement if the density of uncertainty
 belongs to $\mscr{G}$ and the uncertainty is bounded in $\mcal{B}_r$.

An attracting feature of the existing robustness function is that it
relies on very mild assumptions about uncertainty. However, as can
be seen from Theorem 6.1 (in page 856) of \cite{BLT}, the associated
computational complexity can be very high for large uncertainty
dimension.  Another issue of the existing measure is that it can be
very conservative from the perspective of modelling practices. For
illustration of  this point, we consider a conceptual example as
follows.

\begin{figure}
\centering
\includegraphics[height=3.0cm]{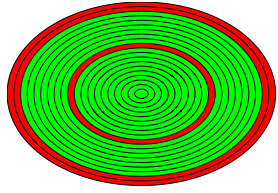}
\caption{Conceptual Example (The robustness requirement is violated
for red layers and is satisfied for green layers. Existing
robustness measure tends to completely ignore uncertainty instances
in the inner layers as $d$ increases. Based on the existing
robustness measure, a very thin bad layer may lead to an unrealistic
judgement that the system has very poor robustness.  However, the
instances in the inner layers are more probably to occur in reality.
Hence, they should have at least equal impact on the evaluation of
system robustness as compared to the instances in the outer layer.)}
\label{fig_Ring}       
\end{figure}

Suppose it is known that the norm of uncertainty $\bs{\vDe}$ cannot exceed $\ga$.  Without loss of generality,
assume $\ga = 1$. That is,  all instances of $\bs{\vDe}$ are included in the bounding set $\mcal{B} = \{ \vDe :
||\vDe|| < 1 \}$.  We partition $\mcal{B}$ as $m$ layers $S_\ell = \{ \vDe: r_{\ell -1} \leq ||\vDe|| < r_\ell
\}, \; \ell = 1, 2, \cd, m$ by radii $r_\ell = \f{\ell}{m}, \; \ell = 0, 1, \cd, m$. From the consideration of
modelling practices, it is reasonable to assume that the density of uncertainty $\bs{\vDe}$ belongs to
$\mscr{F}$.  Hence, for sufficiently large $m$, we have $\Pr \{ \bs{\vDe} \in S_\ell \} \geq \Pr \{ \bs{\vDe}
\in S_{\ell + 1} \}, \; \ell = 1, 2, \cd, m - 1$.  In reality, it is not impossible that not only the outer
layers are ``bad'' and some inner layer is also ``bad''.  Such scenario is described as follows:

The robustness requirement is violated for $\vDe \in S_i$ and for $\vDe \in S_\ell, \; \ell = j, j+1, \cd, m$
where $i$ and $j$ are integers such that $2 \leq i + 1 < j < m$.  See Figure \ref{fig_Ring} for an illustration.
Let $d$ be the dimension of uncertainty space.  By direct computation, we obtain the existing robustness
function as $\underline{\bb{P}}(r) = \inf_{\ro \in [0, r]} \bb{P}(\ro)$ where {\small \[
\bb{P}(\ro) = \bec 1, & \tx{for} \; \ro < r_{i-1};\\
 \f{(i-1)^d }{(m \ro)^d},  & \tx{for} \; r_{i-1} \leq \ro < r_i;\\
\f{(m \ro)^d - i^d + (i-1)^d}{(m \ro)^d},  & \tx{for} \; r_i \leq \ro < r_{j-1};\\
\f{(j - 1)^d - i^d + (i-1)^d}{(m \ro)^d},  & \tx{for} \; r_{j-1} \leq \ro < 1. \eec
\]}
Clearly, $\lim_{d \to \iy} \underline{\bb{P}}(r) = 1$ for $r < r_{i-1}$ and $\lim_{d \to \iy}
\underline{\bb{P}}(r) = 0 $ for $r_{i-1} \leq r < 1$.  This indicates that the existing robustness function
tends to be a discontinuous function as $d$ increases. An undesirable feature of existing measure resulted from
such discontinuity is that a very small variation in the knowledge of the uncertainty bound, $\ga$, may lead to
an opposite evaluation of the system robustness.

For practical systems, large uncertainty instance is less probably while the robustness requirement is more
likely to be violated for larger uncertainty instance. Consequently, unduly conservatism may be introduced if
the uncertainty instances near the surface of uncertainty bounding sets assume a dominant role.  This is indeed
the case for the existing probabilistic robustness measure.  This can be illustrated as follows.  Suppose $\Pr
\{ ||\bs{\vDe}|| < \ga \} = 1$. For the existing measure, the corresponding density of $||\bs{\vDe}||$  of the
sampling distribution that determines $\underline{\bb{P}} (\ga)$ is often times close to {\small
$f_{||\bs{\vDe}||} (r) = d \li ( \f{r}{\ro^*} \ri )^{d-1}$} where $\ro^* = \max \{\ro : \bb{P}(\ro) =
\underline{\bb{P}} (\ga) \}$.  For $\ro \ap \ro^*$, the probability that a sample falls into $\{ \vDe : \ro <
||\vDe|| \leq \ro^* \}$ is $1 - \li ( \f{\ro}{\ro^*} \ri )^d$ which is very close to $1$ when the dimension $d$
is high. This shows that the uncertainty instances near the surface of $\mcal{B}_{\ro^*}$ are dominating in the
evaluation of system robustness.

\section{New Sampling Technique and Robustness Function}

We have shown before that uniform sampling in high dimensional sets suffers from a {\em surface effect}. In the following we introduce a new sampling technique that offsets such effect and we use the modified sampling technique to define the new robustness measure.

\subsubsection{A New Sampling Technique}
To offset the surface effect for uncertainties with radial symmetry we define two independent random variables.
One, $\bs{U}$ is uniformly distributed in the surface of the unit bounding set, $\{ \vDe: ||\vDe|| = 1 \}$, in
the uncertainty space. The second random variable is $\bs{R}$ which is a scalar variable uniformly distributed
over $[0, r]$. Clearly, for a given value of the scalar random variable $\bs{R}$, the uncertainties lay on the
surface of a ball and since $\bs{R}$ is scalar the surface effect is reduced.

\subsubsection{A New
Robustness Function} Now that have established the sampling technique to be used, we define the robustness
measure for the radius $r$ as
\[
\underline{\mscr{P}} (r) = \inf_{ \ro \in (0, \; r]} \mscr{P} (\ro) \;\; \tx{with} \;\; \mscr{P} (r) = \Pr \{
\bb{I}(U R) = 1 \} \]
 where $U$ is a sample from $\bs{U}$ and $R$  a sample from
 $\bs{R}$.  The probabilistic implication of such robustness measure
 can be seen from the
following theorem.

\beT \la{uniform_rad}  For any robustness requirement,
\[
\inf_{f_{\bs{\vDe}} \in \mscr{F}} \Pr \{ \bb{I} (\bs{\vDe}) = 1
\mid ||\bs{\vDe}|| \leq \ga \} = \underline{\mscr{P}} (\ga) \geq
\underline{\bb{P}} (\ga).
\]
 \eeT

See Appendix A for a proof. The intuition behind Theorem
\ref{uniform_rad} is that, in the worst-case, the uncertainty
instances in the inner layers should assume equal importance as that
of uncertainty instances in the outer layers in the evaluation of
system robustness. It should be noted that the density $f_{
||\bs{\vDe}|| } (.)$ can be unbounded and has infinitely many and
arbitrarily distributed discontinuities. An example of unbounded
density is $f_{ ||\bs{\vDe}|| } (\ro) = \f{k-1}{\ro^k}, \; k > 1$.

\begin{figure}
\centering
\includegraphics[height=9cm]{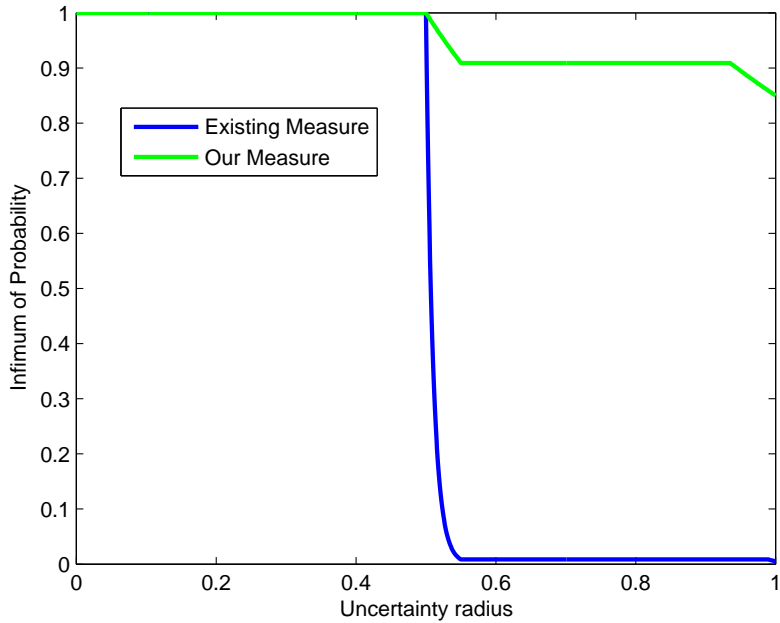}
\caption{Comparison of Robustness Functions ($m = 20, \; i = 11$ and $j = 19$.  The dimension of uncertainty
space is $d = 50$, which is equivalent to a complex block of size $5 \times 5$.)}
\label{fig_Curve4}       
\end{figure}

Now we revisit the conceptual example discussed in Section \ref{3A}.  Our robustness function is
$\underline{\mscr{P}}(r) = \inf_{\ro \in [0, r]} \mscr{P}(\ro)$ where
\[
\mscr{P}(\ro) = \bec 1, & \tx{for} \; \ro < r_{i-1};\\
\f{i-1}{m \ro},  & \tx{for} \; r_{i-1} \leq \ro < r_i;\\
\f{m \ro - 1}{m \ro},  & \tx{for} \; r_i \leq \ro < r_{j-1};\\
\f{j - 2}{m \ro},  & \tx{for} \; r_{j-1} \leq \ro < 1. \eec
\]
As can be seen from Figure \ref{fig_Curve4}, our robustness measure is significantly less conservative than the
existing one.

\sect{Mapping of Robustness Functions}

In this section, we shall demonstrate that there exists a fundamental relationship between our robustness
measure and the existing probabilistic robustness measure. This relationship can be exploited, for example, to reduce the
computational complexity of existing probabilistic robustness measure.

\subsection{Integral Transforms}  The following theorem shows that there exists an integral transform between our proposed robustness function and existing robustness
function.

 \beT
 \la{trans} Define $\phi(r) = \Pr \{ \bb{I}(r U) = 1 \}$
 where $U$ is a random variable uniformly distributed over $\{ \vDe : ||\vDe || = 1 \}$.
 Suppose that the distribution of uncertainty $\bs{\vDe}$ is radially symmetrical
 and that both $f_{||\bs{\vDe}||}(.)$ and $\phi(.)$ are piece-wise
continuous. Then, for any $r > 0$,
 \bee \mscr{P}(r) & = & \f{
\bb{P}(r) } { n } + \f{n-1}{n}  \int_{  0}^1 \bb{P}(r \ro) \; d \ro,\\
 \bb{P}(r) & = & n \; \mscr{P}(r) - n (n-1)
\int_{  0}^1 \mscr{P}(r \ro) \; \ro^{n-1} \; d \ro \eee  where $n$
is the dimension of uncertainty space. \eeT

See Appendix B for a proof. Theorem \ref{trans} shows that once one
of $\mscr{P}(.)$ and $\bb{P}(.)$ is available from Monte Carlo
simulation, the other can be obtained without simulation.

\subsection{Recursive Computation}  For a transform to be useful,
we shall develop efficient method for its computation.  The efficiency can be achieved by recursive computation.
We first discuss the computation of transform from $\mscr{P}(.)$ to $\bb{P}(.)$.

It can be seen that the expression of $\bb{P}(.)$ in terms of
$\mscr{P}(.)$ is not amenable for recursive computation.  By a
change of variable, we rewrite the second equation of Theorem
\ref{trans} as $\bb{P}(r)  = n \; \mscr{P}(r) - \f{ n (n-1) } { r^n}
\int_{  0}^r \mscr{P}(\ro) \; \ro^{n-1} d \ro$.  Clearly, the major
computation is on the integration {\small $I(r) = \int_{  0}^r
\mscr{P}(\ro) \; \ro^{n-1} d \ro$, } which can be computed
recursively because of the relationship {\small $I(r + h) = I(r) +
\int_{ r}^{r+h} \mscr{P}(\ro) \; \ro^{n-1} d \ro$. } Unfortunately,
there will be a numerical problem for computing the product {\small
$\f{ n (n-1) } { r^n} \times I(r)$} in the situation that $n$ is
large and $r < 1$.  For example, {\small $\f{ n (n-1) } { r^n}$} can
be a huge number and cause intolerable numerical error when $n = 36$
and $r = 0.5$. To overcome this problem, we derive the following
recursive relationship {\small \bee \bb{P}(r + h) & = & n \mscr{P}
(r + h) - \li ( \f{r}{r + h} \ri )^n \li [ n \mscr{P}(r) - \bb{P}(r)
\ri ] - \f{ n (n-1) } { (r + h)^n} \int_{  r}^{r + h} \mscr{P}(\ro)
\; \ro^{n-1} d \ro. \eee} Since $\mscr{P}(.)$ can be approximated by
a simple function, we can decompose {\small $\f{ n (n-1) } { (r +
h)^n} \int_{r}^{r + h} \mscr{P}(\ro) \; \ro^{n-1} d \ro$ } as a
summation of integrations of the form {\small $\f{ n (n-1) } {
(r+h)^n} \int_{  \al}^\ba \mscr{P}(\ro) \; \ro^{n-1} d \ro$ } with
$\mscr{P}(\ro) = c, \; \fa \ro \in [\al, \ba]$. Clearly, we have the
explicit formula {\small $\f{ n (n-1) } { (r+h)^n} \int_{ \al}^\ba
\mscr{P}(\ro) \ro^{n-1} d \ro
 =  (n-1) c \li( \f{\al}{r + h} \ri )^n  \li [ \li( \f{\ba}{\al} \ri )^n -  1 \ri ]$.  }

In a similar manner, $\mscr{P}(.)$ can be computed recursively  by relationship \bee \mscr{P}(r + h) & = & \f{
\bb{P}(r + h) } { n } + \f{ r } { r + h } \li [  \mscr{P} (r) - \f{\bb{P}(r)  } { n} \ri ]  + \f{ n - 1 } { n }
\f{1}{r + h} \int_r^{r + h} \bb{P} (\ro) d \ro. \eee

\section{Computational Algorithms and Hierarchial Sample Reuse}

In this section we shall discuss the evaluation of $\mscr{P}(.)$ for
uncertainty radius $\li [\f{a}{\lm}, a \ri ]$ with sample size $N$
and $m$ grid points $\f{a}{\lm} = r_1 < \cd < r_m = a$.
  First, we shall introduce basic subroutines. Second, we present sample
reuse algorithm based on sequential data merging method. Third, we shall demonstrate that the sequential sample
reuse algorithm is impractical and propose hierarchy sample reuse algorithms.

The basic idea of our algorithms is as follows.  Let $U^k, \; k = 1,
\cd, N$ be $N$ i.i.d. samples uniformly generated from $\{\vDe:
||\vDe|| = 1 \}$.  For $i = 1, \cd, m$, we can estimate
$\mscr{P}(r_i)$ as $\f{ \sum_{k = 1}^N \bb{I}(\vDe^{k,i}) } { N }$
with $\vDe^{k,i} = U^k R^{k,i}$ where $R^{k,i}$ is uniformly
distributed over $[0, r_i]$ and is independent of $U^k$ for $k = 1,
\cd, N$.  It should be noted that $R^{k,i}, \; i = 1, \cd, m$ are
not necessarily mutually independent to ensure that $\vDe^{k,i}, \;
k = 1, \cd, N$ are i.i.d samples.  Due to the uniform distribution
of $R^{k,i}$, sample reuse techniques can be employed to save a
substantial amount of computation for the generation of $R^{k,i}, \;
\vDe^{k,i}$ and the evaluation of $\bb{I}(\vDe^{k,i})$ in the
following manner.  Let $k$ be fixed. Let $R$ be a sample uniformly
generated from interval $[0, r_p]$. Then, for any index $j$ such
that $r_j \in [R, r_p]$, we can use $R$ as $R^{k,j}$, $U^k R$ as
$\vDe^{k,j}$, and $\bb{I}(U^k R)$ as $\bb{I}(\vDe^{k,j})$. It can be
shown that the minimum index $j$ can be computed by explicit formula
(\ref{locate}) as \be \la{locate} j = \li \{ \begin{array}{l}
 1 + \max \li (0,  \; \li \lf \f{ (\lm R - a)(m - 1)  } { a( \lm - 1) } \ri \rf \ri ) \; \tx{for uniform gridding;}\\
1 + \max \li (0, \; \li \lf (m-1) \li ( 1 + \f{ \ln \f{R}{a}  } { \ln \lm } \ri ) \ri \rf \ri ) \; \tx{for
geometric gridding} \end{array} \ri. \ee where ``uniform gridding'' means that $r_i - r_{i-1}$ is the same for
$i = 2, \cd, m$ and ``geometric gridding'' means that $\f{r_i}{r_{i-1}}$ is the same for $i = 2, \cd, m$.

For a specific $k$, the sample $U^k$ is referred to as a directional sample and the simulation with sample reuse
techniques to obtain $\bb{I}(\vDe^{k,i}), \; i = 1, \cd, m$ is referred to as ``Radial Sampling''. Clearly,
$\bb{I}(\vDe^{k,i}), \; i = 1, \cd, m$ can be expressed as a matrix $D$ of $3$ columns and random number of rows
such that its $i$-th row $[D_{i1}, \; D_{i2}, \; D_{i3}]$ means  that
\[
\bb{I}(\vDe^{k,j}) = \bec 1 & \tx{if} \; D_{i3} = 1;\\
0 & \tx{if} \; D_{i3} = 0 \eec
\]
for $D_{i1} \leq j \leq D_{i2}$.   The algorithm of ``Radial Sampling'' is formally described in Section
\ref{Radial Sampling}.

The process of obtaining the summation $\sum_{k =1}^N \bb{I}(\vDe^{k,j}), \; i = 1, \cd, m$ is accomplished by
the subroutine ``Merging'', which is described in Section \ref{Merging}.

\subsection{Radial Sampling} \la{Radial Sampling}

For a directional sample $U$, the goal of radial sampling is to create a matrix $D$.  The input of the
subroutine ``Radial Sampling'' is $U, \lm, a,  m$ and the corresponding output is $D = \mrm{RS} (U, \lm, a, m)$.
The algorithm is presented as follows.

 \hrulefill

 \bei

\item Let $p \leftarrow m$ and do the following.
   \bei
      \item Generate a sample $R$ uniformly from $[0, r_{p}]$.

     \item Let $\vDe \leftarrow U R$ and evaluate
     $\bb{I}(\vDe)$.

     \item Determine the smallest index $j$ such that $r_{j} \geq R$ by (\ref{locate}).

     \item Let $D \leftarrow [j, \; p, \; \bb{I}(\vDe)]$ and $s \leftarrow \bb{I}(\vDe)$.

     \item Let $p \leftarrow j - 1$.
     \eei

\item While $p > 0$ do the following.

     \bei
      \item Generate a sample $R$ uniformly from $[0, r_p]$.

       \item Let $\vDe \leftarrow U R$ and evaluate $\bb{I}(\vDe)$.

       \item Determine the smallest index $j$ such that $r_{j} \geq R$ by (\ref{locate}).

       \item If $\bb{I}(\vDe) \neq s$, add $[j, \; p, \; \bb{I}(\vDe)]$ to $D$ as the first row and let
         $s \leftarrow \bb{I}(\vDe)$.  Otherwise, update the first element of the first row of $D$ as $j$.

       \item Let $p \leftarrow j - 1$.

\eei

\item Return $D$ as the outcome of radial sampling.

\eei

\hrulefill

\subsection{Merging} \la{Merging}

The operation of merging involves two matrices $D$ and $H$. Matrix $D$ defines a segmented function $f_D(.)$
over domain $\{1, \cd, m\}$ in the sense that, for the $j$-th row of $D$, $f_D(i) = D_{j3}$ for any $i$ such
that $D_{j1} \leq i \leq D_{j2}$.  Similarly, matrix $H$ defines a segmented function $f_H(.)$ over domain $\{1,
\cd, m\}$ in the sense that, for the $j$-th row of $H$, $f_H(i) = H_{j3}$ for any $i$ such that $H_{j1} \leq i
\leq H_{j2}$.  For input matrices $D$ and $H$, the merging operation finds $M = \mrm{Merge}(D,H)$ such that
\[
f_M(i) = f_D(i) + f_H(i), \qqu i = 1, \cd, m \]
 where $f_M(.)$ is a segmented function $f_M(.)$ over domain $\{1,
\cd, m\}$ in the sense that,  for the $j$-th row of $M$, $f_M(i) = M_{j3}$ for any $i$ such that $M_{j1} \leq i
\leq M_{j2}$.

\subsection{Sequential Sample Reuse Algorithm (SSRA)}

The sequential algorithm derives its name from the sequential nature
of the data merging process.  The input variable is $N, \lm, a,  m$
and the output is a matrix $H$ of random number of rows and $3$
columns. The main algorithm is presented as follows.

 \hrulefill

\bei

\item Let $k \leftarrow 1$ and do the following.

\bei
\item Generate a directional sample $U$.

\item Perform radial sampling and let $D \leftarrow \mrm{RS} (U, \lm, a, m)$.

\item Let $H \leftarrow D$.

\eei

\item While $k < N$ do the following.
     \bei

     \item Generate a directional sample $U$.

     \item Perform radial sampling and let $D \leftarrow \mrm{RS} (U, \lm, a, m)$.

     \item Perform merging and let $H \leftarrow \mrm{Merge}(D, H)$.

     \item Let $k \leftarrow k + 1$.

     \eei
\item Return $H$.

\eei

\hrulefill

\bsk

Once we have $H$ from the execution of SSRA, we can estimate
$\mscr{P}(r_i)$ as $\f{\sum_{k = 1}^N \bb{I}(\vDe^{k,i})}{N} =
\f{f_H(i)}{N}, \; i = 1, \cd, m$.

\subsection{Hierarchy Sample Reuse Algorithm (HSRA)}

A major problem with the sequential algorithm is that the computational effort devoted to merging becomes an
enormous burden as the sample size $N$ becomes large.

\begin{figure} \centering
\includegraphics[height=9cm]{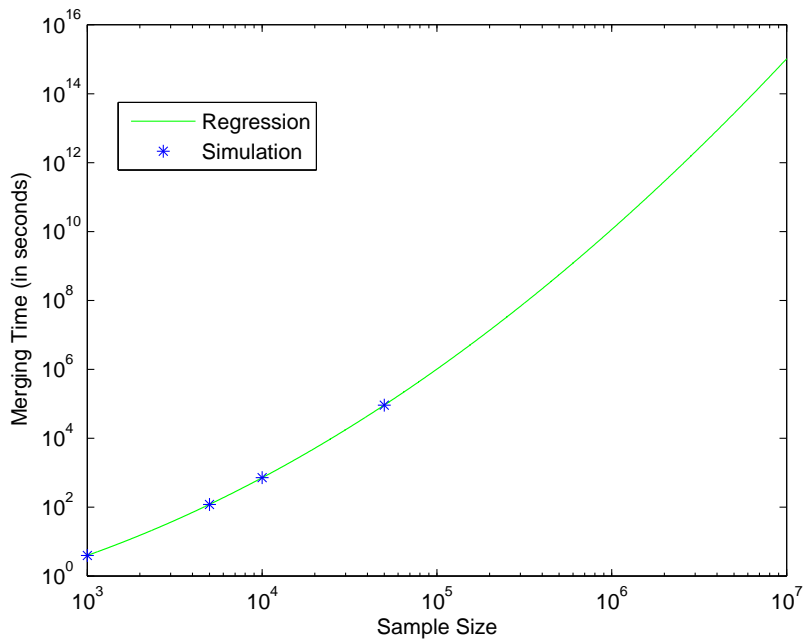}
\caption{Merging Time}
\label{fig_merging_time}       
\end{figure}

The merging time for $N = 1000, \; 5,000, \; 10,000$ and $50,000$ are respectively $4, \; 120, \; 722$ and
$92119$ seconds, which is obtained by simulation on a PC of $1024$M RAM and $3.2$G CPU.  As can be seen from
Figure \ref{fig_merging_time}, the merging time required for $N = 10^5, \; 10^6$ and $N = 5 \times 10^6$ is
predicted respectively as, $12$ days, $366$ years, and $9 \times 10^5$ years, by fitting the simulation data
into a quadratic function (in log scale) based on regression techniques.  For a better understanding of the
complexity issue, a theoretical analysis of the computational complexity of data merging is as follows.

From the merging process, it can be seen that the computational
complexity of merging two matrices can be quantified by the sum of
the numbers of the rows of the two input matrices.  Thus, it
suffices to study how the number of rows is growing when matrices
$D^k = \mrm{RS} (U^k, \lm, a, m), \; k = 1, \cd, N$ are sequentially
merged.

Note that the average numbers of rows for all $D^k$ are identical.
Let this average be $L$. To merge $D^1$ with $D^2$, the required
computation is $2L$.  The computation to merge the outcome with
$D^3$ is $3L$. The computation for all steps of merging forms a
series, $2L, \; 3L, \; \cd, NL$, of constant increment $L$. Hence,
the total number of computation is {\small $\f{L(N + 2) (N-1)}{2}$.
}  This can be a huge number because $N$ is usually large.

To overcome the difficulty of sequential algorithm, we propose a
merging method of hierarchy structure.  We first introduce a
subroutine called {\it successive binary merging} for $N = 2^p$ data
matrices as follows.

\begin{figure}
\centering
\includegraphics[height=5cm]{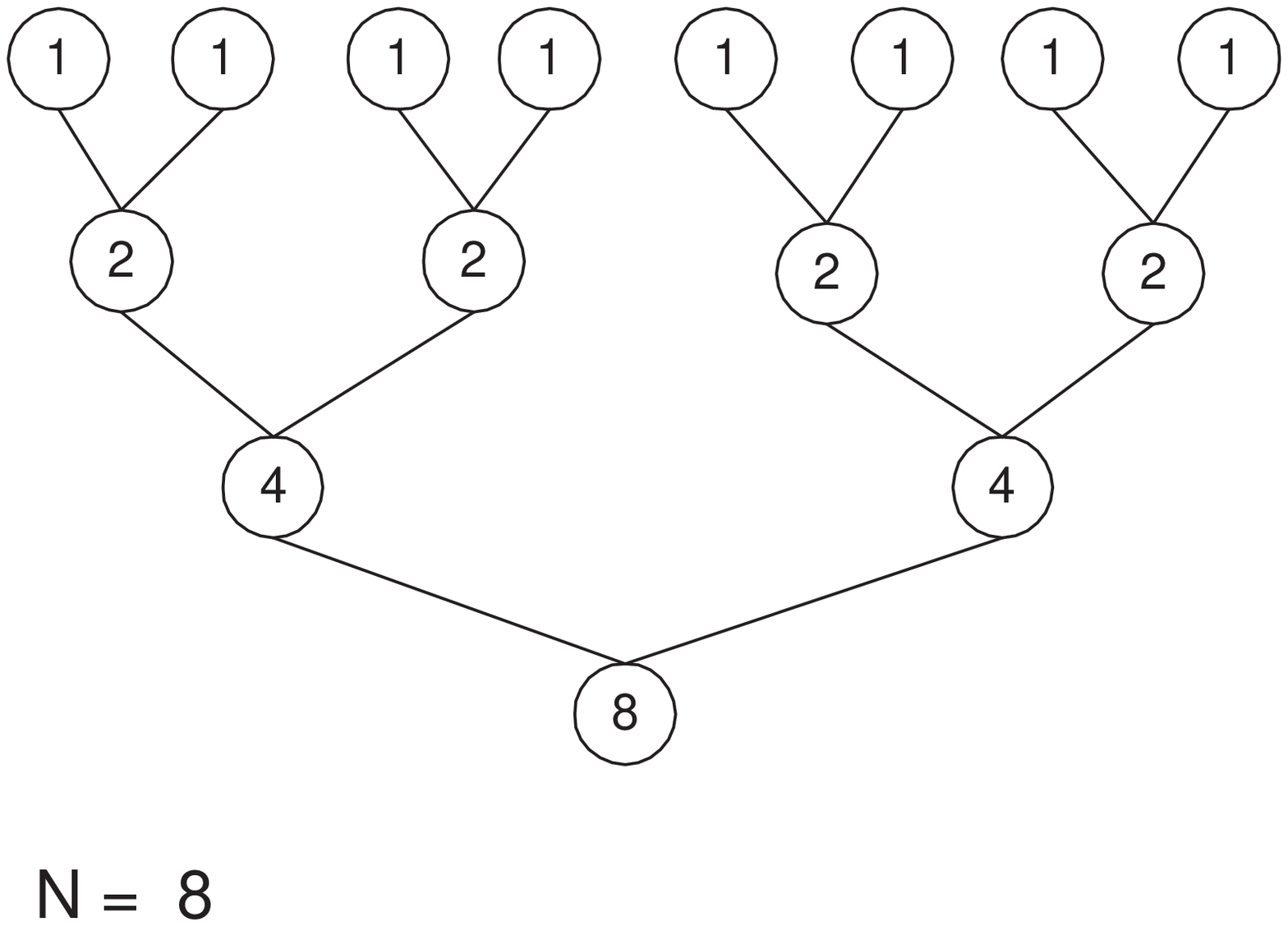}
\caption{Illustration of Successive Binary Merging with $N = 8$. }
\label{SBM}       
\end{figure}

Divide these $N$ matrices $D^1, \cd, D^N$ into $\f{N}{2}$ groups so
that each group has two matrices. After merging each group, we have
$\f{N}{2}$ matrices. Repeating the operations of dividing and
merging, we obtain a matrix in the final stage. This process can be
associated with a binary tree as illustrated by Figure \ref{SBM}.

For the general case that $N$ is not a power of $2$, we decompose
$N$ as a summation of numbers which are powers of $2$. For example,
for $N = 1000$, we have $N = 512 + 256 + 128 + 64 + 32 + 8$.  Such
decomposition corresponds to the decimal-to-binary conversion.  In
general, for $N = \sum_{\ell = 1}^\tau N_\ell$ with $N_\ell =
2^{p_\ell}$ and $N_1 < N_2 <\cd < N_\tau$, the merging can be
performed as follows.

\hrulefill

\bei

\item Let $\ell  \leftarrow 1$. Applying successive binary merging to $N_1$ to create data matrix $M_1$.
Let $H  \leftarrow M_1$.

\item While $\ell < \tau$ do the following.

    \bei
    \item Applying successive binary merging to $N_\ell$ to create data matrix $M_\ell$.
    \item Let $H \leftarrow \mrm{Merge}(H, M_\ell)$.
    \item Let $\ell  \leftarrow \ell + 1$.
    \eei

\eei

\hrulefill

\begin{figure}
\centering
\includegraphics[height=6cm]{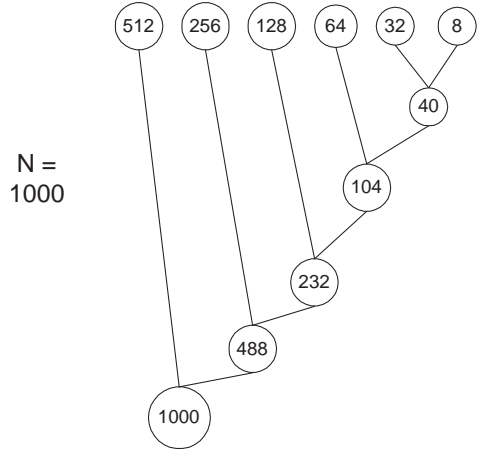}
\caption{Merging with $N = 1000$. }
\label{DM}       
\end{figure}

The merging for $N = 1000$ is shown by Figure \ref{DM}.

The complexity of such hierarchy can be analyzed as follows. For
successive binary merging with $N = 2^p$, the computation is $p
\times N L$.  For $N = \sum_{\ell = 1}^\tau N_\ell$, the computation
is bounded by $L \sum_{\ell = 1}^\tau N_\ell \log_2(N_\ell) + L
\sum_{\ell = 1}^\tau (\tau - \ell + 1) N_\ell - L N_1$.  Therefore,
the computation is reduced from the sequential algorithm by a factor
of {\small $\Up = \f{(N+2) (N-1)} { 2 \li [  \sum_{\ell = 1}^\tau
N_\ell \log_2(N_\ell) + \sum_{\ell = 1}^\tau (\tau - \ell + 1)
N_\ell - N_1\ri ] }$. } Specially, for $N = 2^p$, we have $\Up =
\f{(N+2) (N-1)} { 2 \log_2 (N) } > \f{N}{2 \log_2(N) }$, which is
usually a very large number.

\section{Computational Complexity}

In this section, we discuss the computational complexity for the
evaluation of $\mscr{P}(.)$ over uncertainty radius interval
$(\f{a}{\lm}, a]$.  For practical designs, the robustness
requirement is guaranteed for the nominal model. Hence, $\mscr{P}
(\ro) = 1$ for small $\ro$, and we have $\inf_{ \ro \in (0, a]}
\mscr{P} (\ro) = \inf_{ \ro \in (\f{a}{\lm}, \ga]} \mscr{P} (\ro)$
for a sufficiently large $\lm$. A direct Monte Carlo simulation
method is to partition the interval $(\f{a}{\lm},  a]$ by $m$ grid
points $\f{a}{\lm} = r_1 < \cd < r_m = a$ and estimate $\mscr{P}
(r_i)$ by $N$ i.i.d. Monte Carlo simulations. The estimate of
$\inf_{ \ro \in (\f{a}{\lm},  a]} \mscr{P} (\ro)$ is obtained by
taking the minimum of the results for the $m$ grid points. Such
direct method requires $m N$ simulations. As $m$ gets large, the
computing time and the memory complexity becomes a challenging
problem. Fortunately, by employing our hierarchy sample reuse
algorithms, {\it the computational complexity is absolutely bounded
and very low for arbitrarily dense griding and arbitrarily large
dimension of uncertainty}.

For quantifying the computational complexity, we define the {\it equivalent number of grid points},
$m_{\mrm{eq}}$ as the ratio \[ m_{\mrm{eq}} = \f{\tx{ Average total number of simulations} } { N }.
\]

We shall interpolate the value of $\mscr{P}(r)$ for $r \in [r_i, r_{i+1}]$ as {\small
\[
\mscr{P}^*(r) = \f{ (r - r_i) \; \mscr{P} (r_{i+1}) + (r_{i+1} - r ) \; \mscr{P} (r_{i}) } { r_{i+1} - r_i }.
\]
}

For a uniform gridding, we have
 \beT \la{grid_uni}  Let $\ep \in (0,1)$ and $m = 2 + \li \lf \f{ 2(\lm -1)} { \ep  } \ri \rf$. Let
 {\small $r_i = \f{a}{\lm} + \f{ (i-1) \li ( a - \f{a}{\lm} \ri ) } { m - 1 }$} for $i = 1, \cdots, m$. Then,
\[
 |\mscr{P} ( r ) -  \mscr{P}^* ( r)| < \ep, \qu \fa r \in [r_i, r_{i+1}]
 \]
  for $i = 1, \cd, m-1$. Moreover,
 {\small $m_{\mrm{eq}} (\epsilon) = m - \sum_{i=1}^{m-1}
 \li (  1 - \f{1} { \f{m-1} {\lm-1} + i } \ri ) < 1 +  \ln \lm$} for any $\ep \in (0,1)$. \eeT

\bsk See Appendix C for a proof.
 For a geometric gridding, we have
 \beT \la{Grid_geometric} Let $\ep \in (0,1)$ and
 {\small $m = 2 + \li \lf \f{ \ln  \lm  } { \ln \li ( 1 + \f{\ep}{2} \ri ) } \ri \rf$}.
 Let $r_i = a \li (  \f{1}{\lm} \ri )^ {\f{m - i} {m-1} }$ for $i = 1, \cdots, m$. Then,
\[
 |\mscr{P} ( r ) -  \mscr{P}^* ( r)| < \ep, \qu \fa r \in [r_i, r_{i+1}]
 \]
 for $i = 1,\cd,m - 1$.  Moreover,
  {\small $m_{\mrm{eq}}(\epsilon) =  1 + \li ( m - 1 \ri ) \; \li [ 1 - \left( \frac{1}{\lm} \right)^{ \f{1}{
m - 1  } } \ri ] < 1 +  \ln \lm $} for any $\ep \in (0,1)$.

\eeT

See Appendix C for a proof. For completeness, we note that,  for
arbitrarily large $m$, the memory complexity is also absolutely
bounded and independent of uncertainty dimension.

\bsk

 To compare the computational complexity of our probabilistic
measure with that of \cite{BLT}, we recall Theorem 6.1 of
\cite{BLT}, which states that if \be \la{Bar_La_Te} m \geq 1 + \f{ 2
(\lm -1) d } { \ep } \ee then $|\bb{P} ( r ) - \bb{P}( r_i)| < \ep
\qu \fa r \in [r_i, r_{i+1}]$ for $i = 1, \cd, m-1$. This bound
shows that, for fixed error $\ep$, the complexity is polynomial.
From another perspective, it also shows that the number of grid
points and computational complexity tend to infinity as the
tolerance tends to zero.  The computational complexity can be
reduced by the sample reuse techniques of \cite{C0}. It is recently
shown in \cite{Chen_SIAM} that the equivalent number of grid points
is bounded by $1 + d \ln \lm$ (see Appendix C for a proof).  In
applications, $d$ can be very large. For example, the dimension $d$
is $2 n^2$ for a complex block of size $n\times n$. Since the
complexity of computing $\mscr{P}(.)$ is independent of dimension
$d$, the integral transform can be applied to obtain $\bb{P}(.)$
from $\mscr{P}(.)$ and thus significantly reduced the computational
complexity.

\section{Examples}

In this section, we shall demonstrate the power of our techniques by examples.  By the definition of  the
indicator function $\bb{I}(.)$, for $N$ i.i.d. samples $\vDe_1, \cd, \vDe_N$ generated from $\mcal{B}_r$,
\[
\bb{I} (\vDe_i) = \bec 1 & \tx{if the robustness requirement is satisfied for $\vDe_i$};\\
0 & \tx{otherwise}. \eec
\]
Specially, for the robustness stability problem in the $M-\vDe$ setup with $M(s) = C(sI - A)^{-1} B$,
\[
\bb{I} (\vDe_i) = \bec 1 & \tx{if} \; A + B \vDe_i C \; \tx{is stable};\\
0 & \tx{otherwise}. \eec
\]
Of course, the $N$ samples are obtained by the HSRA.  A minimum
variance unbiased estimator of $\mscr{P}(r)$ is taken as
$\wh{\mscr{P}}(r) = \f{\sum_{i=1}^N \bb{I} (\vDe_i)}{N}$.  Since
$\bb{I} (\vDe_i), \; i = 1, \cd, N$ are i.i.d. Bernoulli random
variables with a success probability $\mscr{P}(r)$, the Chernoff
bound \cite{Chernoff} asserts that, for any $\vep, \; \de \in
(0,1)$, $\Pr \li \{ \li | \wh{\mscr{P}}(r) - \mscr{P}(r) \ri | <
\vep \ri \}
> 1 - \de$
if the sample size $N > \f{ \ln \f{2}{\de} } { 2 \vep^2 }$.

In all examples, we first apply our previous method in \cite{C2} to obtain an estimate of the probabilistic
margin with a risk probability $\al = 0.05$ (Roughly speaking, we are only interested in the curve of robustness
function above $1 - \al = 0.95$). Then, we evaluate the robustness function $\underline{\mscr{P}}(r)$ for $r \in
[\f{a}{e}, a]$ by our hierarchy sample reuse algorithms.  The existing robustness measure is computed from our
measure by the integral transform.  The algorithms are implemented in MATLAB and all programs are executed on a
PC of $1024$M RAM and $3.2$G CPU.

We first consider the case that the uncertainty is of a single
block.  A typical robustness problem is to determine the robustness
margin which is specified as the maximum size of uncertainty under
the condition that all poles of the closed-loop system are
restricted in a certain domain $\bb{C}_g$.  For single blocked
uncertainty, there exists formulas for computation of the robustness
margin in a $M - \De$ setup with $M(s) = C(sI - A)^{-1} B$ (see,
e.g., \cite{ZDG} for illustration). For complex uncertainty, the
robustness margin is \bee r_{\bb{C}} & = & \inf \{
\overline{\si}(\vDe) : \vDe \in \bb{C}^{m \times p} \; \mrm{and \;
all\; eigenvalues \; of} \; A + B \vDe C \; \mrm{are \;  in} \;
\bb{C}_g \} =  \f{ 1 }{ \sup_{s \in \pa \bb{C}_g } \overline{\si} [C
(s I - A)^{-1} B] } \eee where $\pa \bb{C}_g$ denotes the boundary
of domain $\bb{C}_g$.  This formula was essentially obtained by
Doyle and Stein \cite{Doyle}. For real uncertainty, the robustness
margin is  \bee r_{\bb{R}} & = & \inf \{ \overline{\si}(\vDe) : \vDe
\in \bb{R}^{m \times p} \; \mrm{and \; all\; eigenvalues \; of} \; A
+ B \vDe C \; \mrm{are \; in} \; \bb{C}_g
\}\\
& = & \f{ 1 } { \sup_{s \in \pa \bb{C}_g } \inf_{\ga \in (0,1]}
\si_2 \li ( \bem \Re (M) & - \ga \;
\Im (M) \\
\ga^{-1} \; \Im (M) & \Re (M) \eem  \ri ) }
 \eee
 where the function to be minimized is a unimodal function on
 $(0,1]$.  This formula was established by Qiu and his coworkers \cite{Qiu}.

\begin{figure}[htbp]
\centerline{\psfig{figure=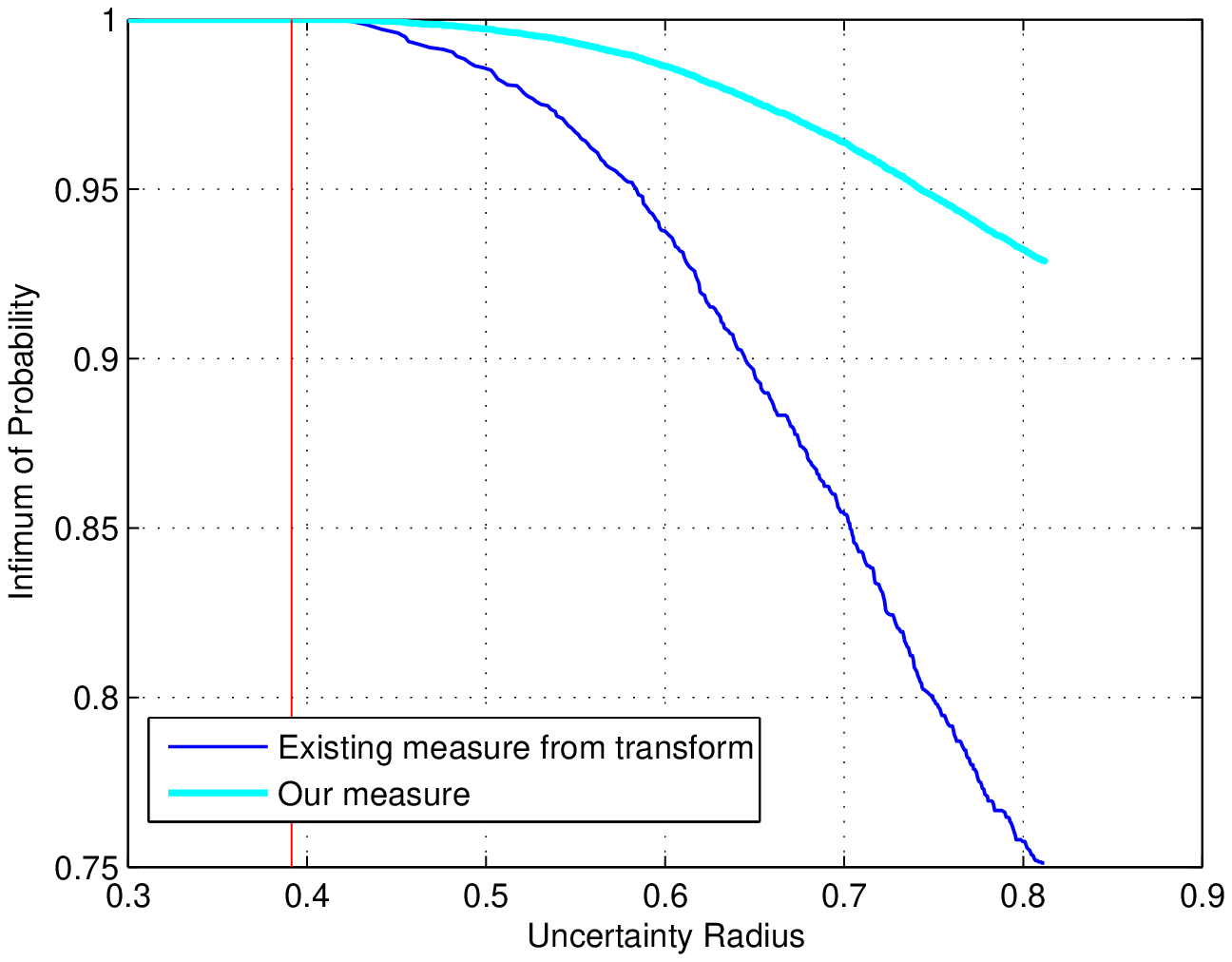, height=3.8in, width=5in }} \caption{Robustness Functions
(Sample Size $N = 26482$). The vertical line marks the deterministic robustness margin.}
\label{fig_Qiu_Exam2_complex}
\end{figure}

\begin{figure}[htbp]
\centerline{\psfig{figure=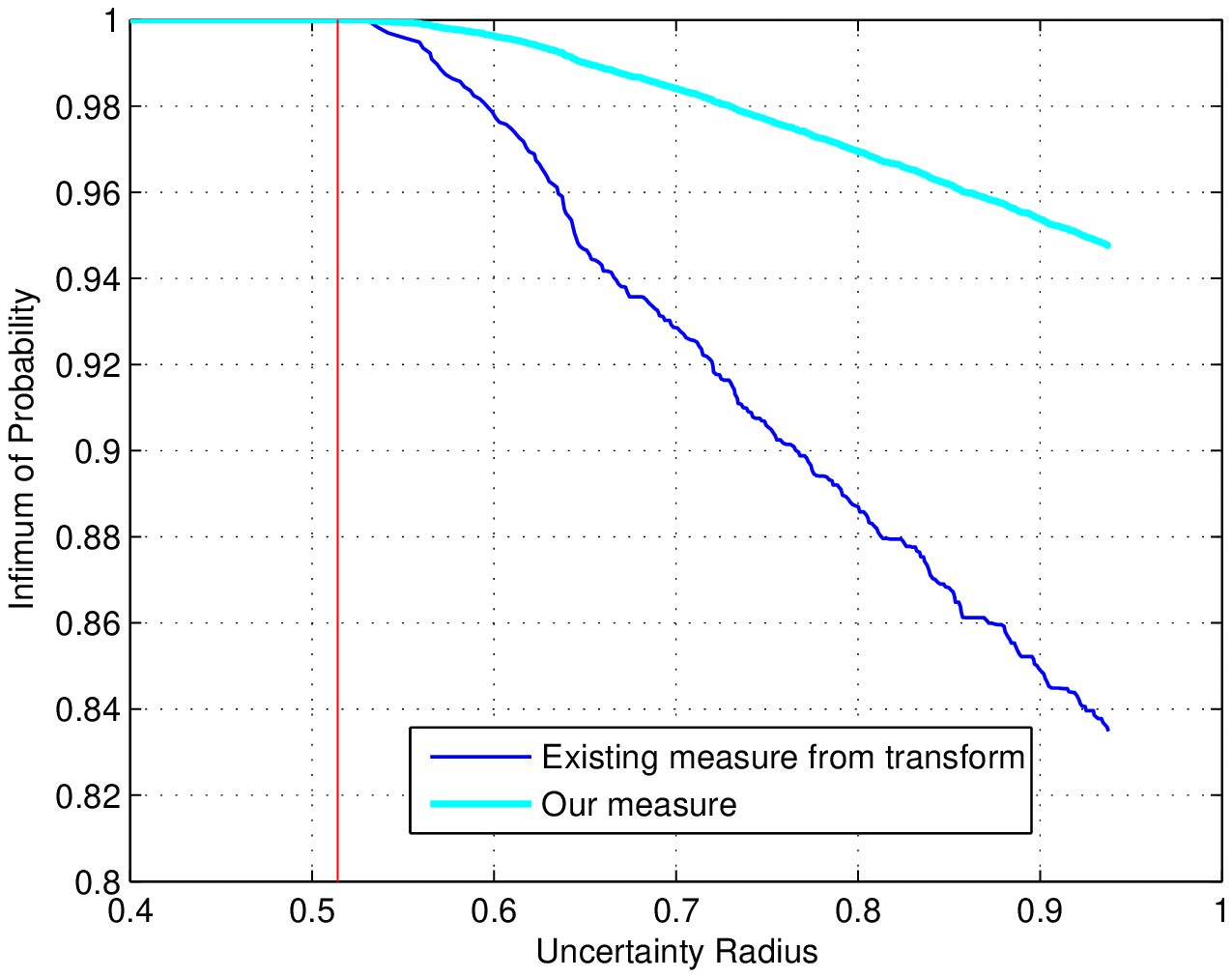, height=3.8in, width=5in }} \caption{Robustness Functions (Sample
Size $N = 26482$). The vertical line marks the deterministic robustness margin.} \label{fig_Qiu_Exam2_real}
\end{figure}

To compare the power of our randomized algorithms with that of these
formulas, we revisit two examples of \cite{Qiu}.  In example $2$ of
\cite{Qiu}, the domain $\bb{C}_g$ is defined as $\bb{C}_g = \{s \in
\bb{C} : \Re(s) < 0 \}$.  The data of matrices $A, \; B,\; C$ can be
found in page $889$ and is thus omitted here.  The robustness
margins for the complex and real uncertainty are obtained,
respectively, as $r_{\bb{C}} = 0.3914$ and $r_{\bb{R}} = 0.5141$.
The robustness functions are shown in Figures
\ref{fig_Qiu_Exam2_complex} and \ref{fig_Qiu_Exam2_real} for the
cases of complex and real uncertainty respectively.  It can be seen
that our randomized algorithms can provide useful information for
the system robustness beyond the deterministic robustness margin.
Specially, the deterministic robustness margin can be estimated from
both types of robustness functions.  Moreover, it can be seen that
our robustness measure is significantly less conservative than the
existing robustness measure.

\begin{figure}[htbp]
\centerline{\psfig{figure=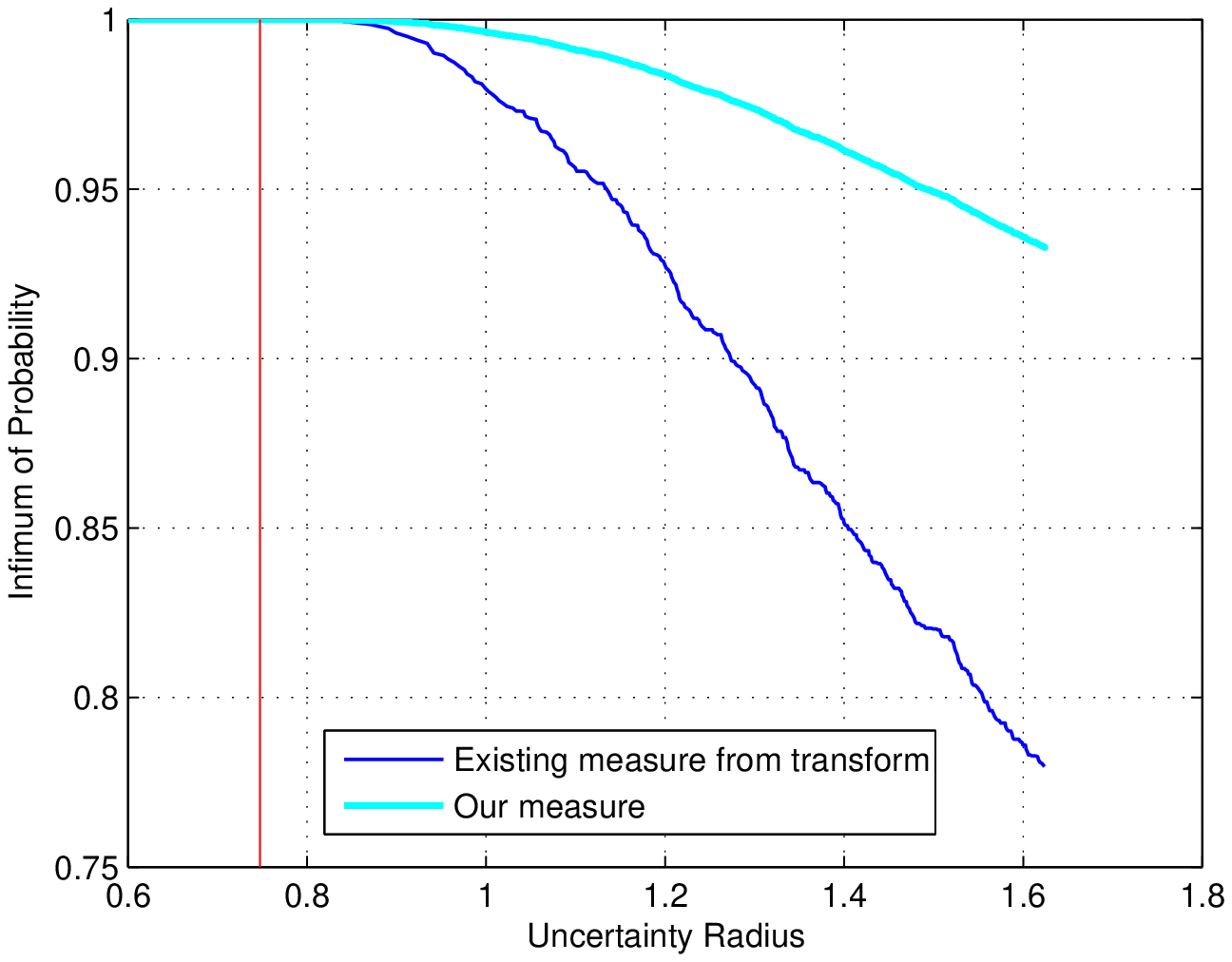, height=3.8in, width=5in }} \caption{Robustness Functions
(Sample Size $N = 26482$). The vertical line marks the deterministic robustness margin.}
\label{fig_Qiu_Exam3_complex}
\end{figure}

\begin{figure}[htbp]
\centerline{\psfig{figure=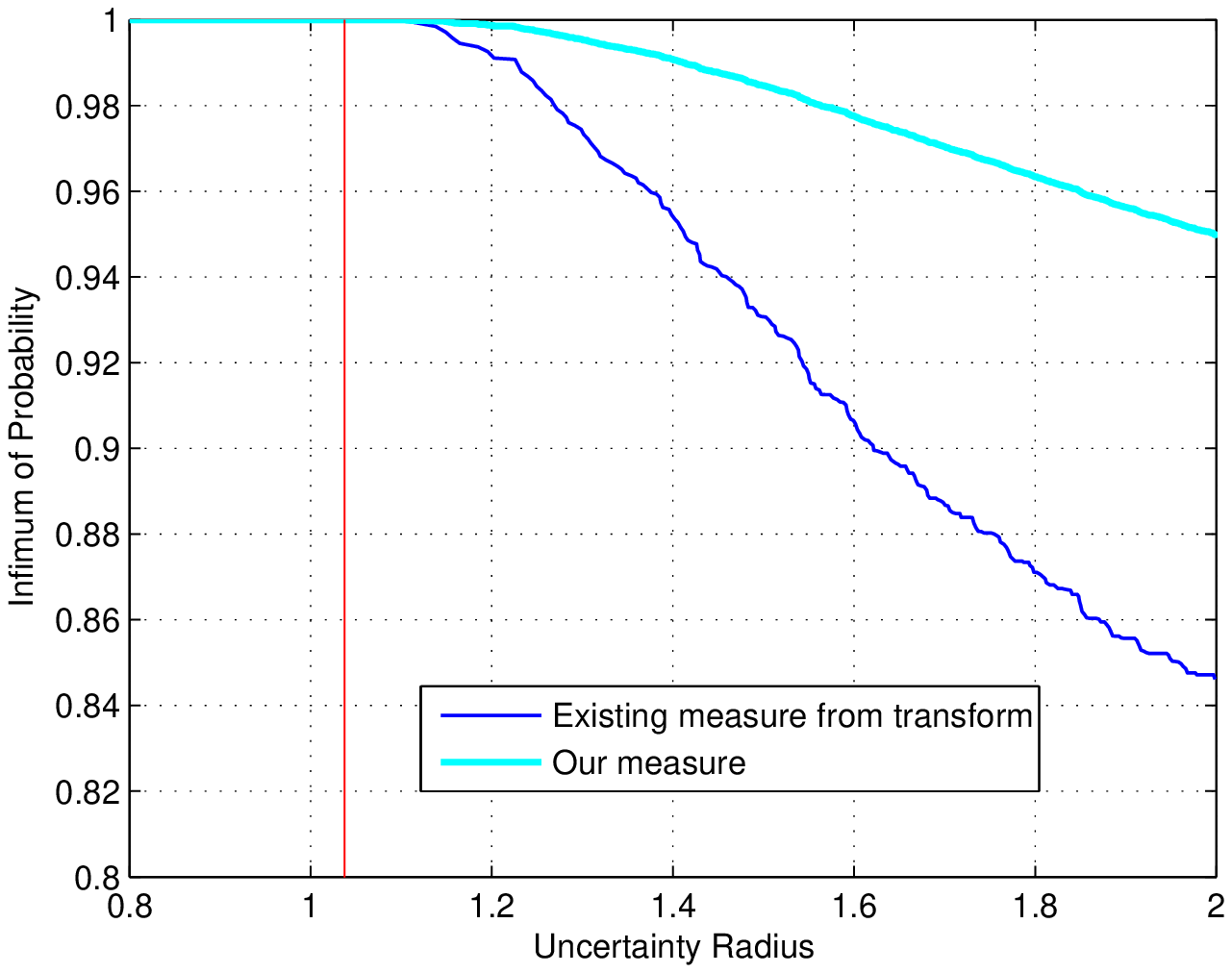, height=3.8in, width=5in }} \caption{Robustness Functions (Sample
Size $N = 26482$). The vertical line marks the deterministic robustness margin.} \label{fig_Qiu_Exam3_real}
\end{figure}

In example $3$ of \cite{Qiu}, the domain $\bb{C}_g$ is defined as
$\bb{C}_g = \{s \in \bb{C} : |s| < 1 \}$ and the data of matrices
$A, \; B,\; C$ are given in page $889$.  The robustness margins for
the complex and real uncertainty are obtained as $r_{\bb{C}} =
0.7472$ and $r_{\bb{R}} = 1.0374$ respectively.  The robustness
functions are shown in Figures \ref{fig_Qiu_Exam3_complex} and
\ref{fig_Qiu_Exam3_real} for the cases of complex and real
uncertainty respectively.

\begin{figure}[htbp]
\centerline{\psfig{figure=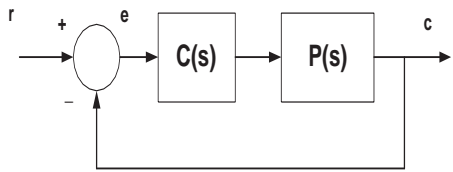, height=1.0in, width=2.7in }} \caption{Uncertain System }
\label{fig_Drawing1}
\end{figure}

We now consider the stability margin problem where the uncertainty
consists of multiple blocks.  A particularly important special case
is that the uncertainty is real parameters.  When the number of
uncertainty blocks is more than one, the formulas of \cite{Doyle}
and \cite{Qiu} are not applicable and the branch and techniques are
needed. We explore the application of our HSRA for the stability
margin problem studied in \cite{GS} by a deterministic approach. The
system considered in \cite{GS} is represented by
Figure~\ref{fig_Drawing1}.  The compensator is
$C(s)=\frac{s+2}{s+10}$ and the plant is {\small
$P(s)=\frac{800(1+0.1\delta_1)}{s(s+4+0.2\delta_2)(s+6+0.3\delta_3)}$
} with parametric uncertainty $\vDe =[\delta_1,\delta_2,\delta_3]$.

\begin{figure}[htbp]
\centerline{\psfig{figure=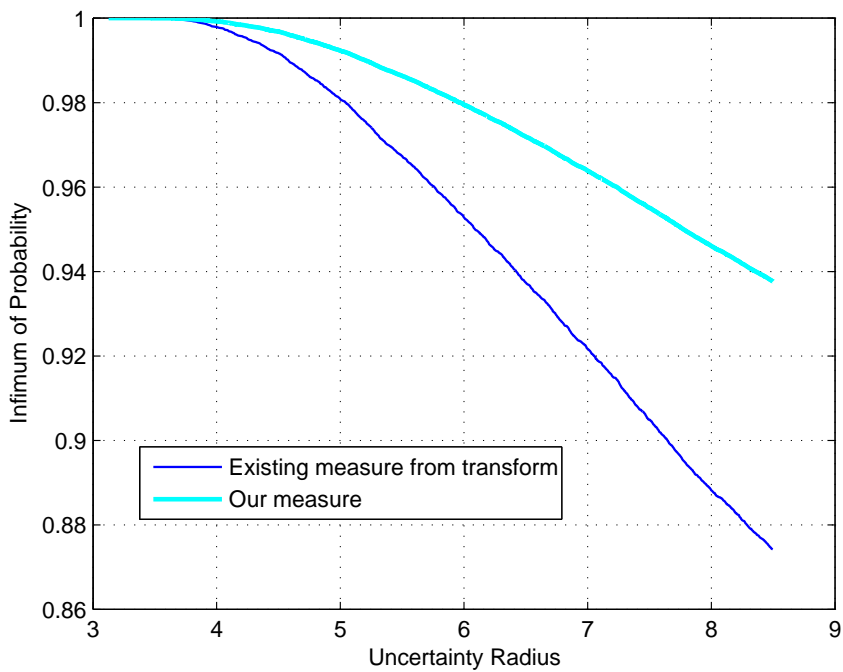, height=3.8in, width=5in }} \caption{Robustness Functions (Sample
size $N = 119,830$.)} \label{fig_Sofonov_system}
\end{figure}

The deterministic robustness margin is found to be $3.44$ by a
branch and bound technique (see, page 163 of \cite{GS}).  The
robustness functions are shown in Figure \ref{fig_Sofonov_system},
which provides more insight for the system robustness than the
deterministic robustness margin.

\begin{figure}[htbp]
\centerline{\psfig{figure=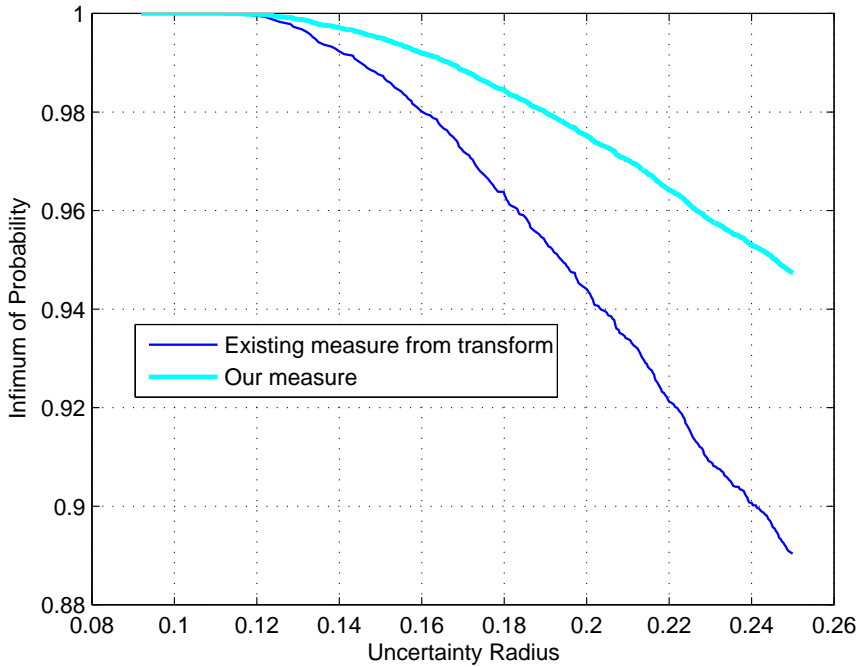, height=3.8in, width=5in }} \caption{Robustness Functions
(Sample size $N = 26482$.)} \label{fig_Sofonov_system_time}
\end{figure}

We now consider the robustness problem involving time-domain specifications for the same system shown by Figure
\ref{fig_Drawing1}. The robustness requirement is that the rise time and settling time should be no more than
$0.25$ and $3.5$ seconds respectively and the overshoot should be no more than $70 \%$ under the condition that
the closed-loop system is stable. It is well-known that this type of problems are, in general, intractable by
the deterministic approach. However, our HSRA can readily provided insightful solution.  The robustness
functions are shown in Figure \ref{fig_Sofonov_system_time}.

Now we present more extensive numerical experiments for testing the
efficiency of our hierarchy sample reuse algorithms. We consider the
robust stability of a system of transfer function $H(s) = C (s I -
A)^{-1} B + D$ with uncertain matrix $A = - 10 \; I_{k \times k} +
\sum_{\ell=1}^d q_\ell \; \sq{\ell} \; W$ where $I_{k \times k}$ is
a $k$ by $k$ identity matrix, $d = k^2$ is the dimension of
uncertainty and $W$ is a matrix with all elements equal to $1$. This
is a special case of multiple blocks of real uncertainty.  Although
this may not be a realistic system, it can be representative for
realistic systems in the respect of computational complexity.

\begin{figure}[htbp]
\centerline{\psfig{figure=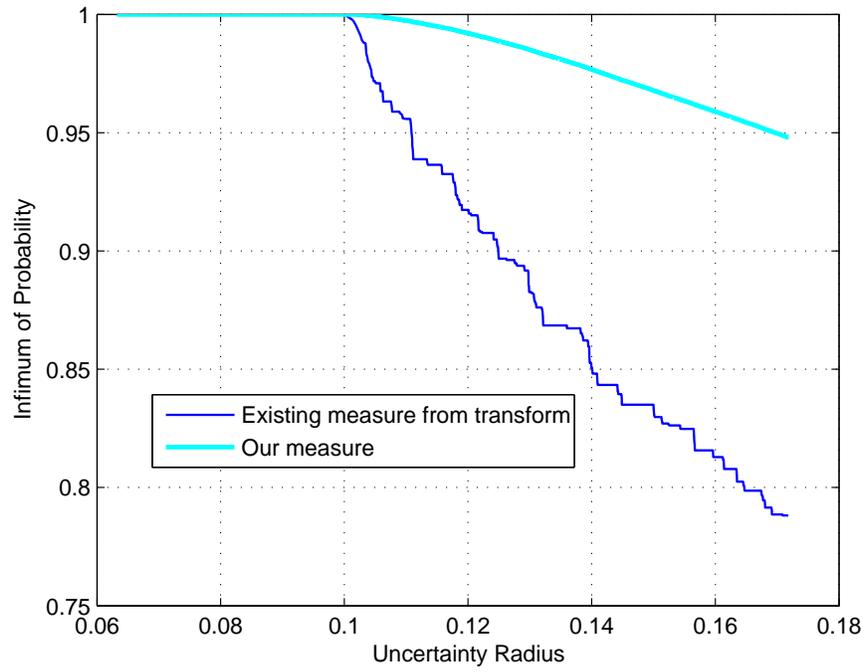, height=3.8in, width=5in }} \caption{Robustness Functions
(Dimension $d = 100$. Sample size $N = 119,830$.)} \label{fig_RC_al005dim100}
\end{figure}

\begin{figure}[htbp]
\centerline{\psfig{figure=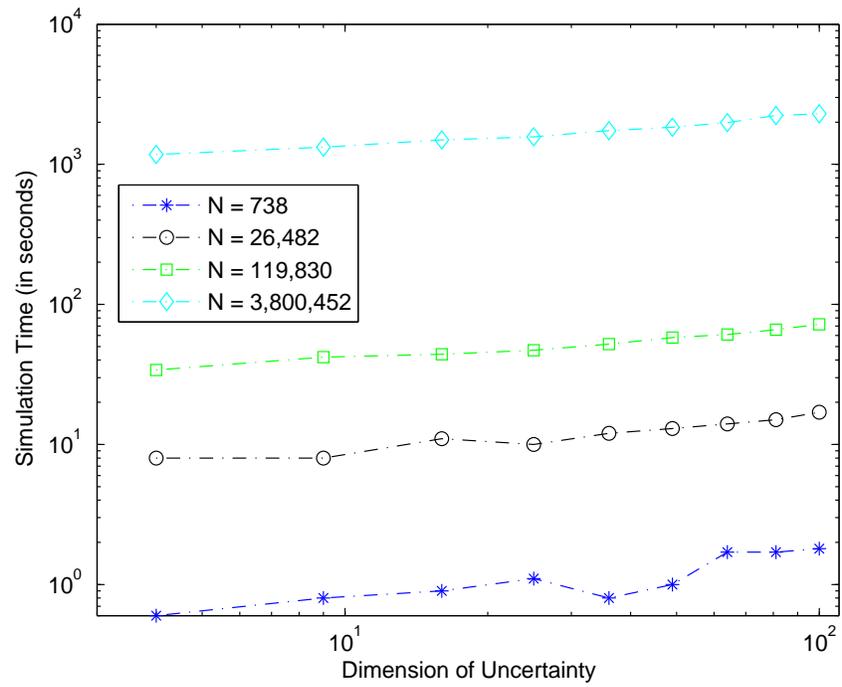, height=3.8in, width=5in }} \caption{Simulation Time}
\label{fig_Simulation_Time}
\end{figure}

When the size of matrix $A$ increases from $2$ to $10$, the dimension of uncertainty increases from $4$ to
$100$.  The robustness functions for the case that $A$ is of size $10 \times 10$ is shown in Figure
\ref{fig_RC_al005dim100}.  The computing time is shown in Figure \ref{fig_Simulation_Time} for various problem
sizes.  The sample size is chosen by the Chernoff bound {\small $N = \li \lceil \f{ \ln \f{2}{\de} } { 2 \vep^2
} \ri \rceil$} as $738, \; 26482, \; 119830, \; 3800452$ corresponding to $\vep = \de = 0.05, \; 0.01, \; 0.005,
\; 0.001$ respectively.

Traditionally, it is widely believed that the classical deterministic robustness analysis are usually more
efficient than randomized algorithms. However, as can be seen from Figure \ref{fig_Simulation_Time}, our
numerical experiments indicates that, if one is willing to accept our probabilistic robustness measure, the
robustness analysis via hierarchy sample reuse algorithms can be generally far more efficient.

\sect{Conclusion}

In this paper, we develop a new statistical approach for robustness
analysis which requires an extremely low complexity that is
independent of the dimension of uncertainty space.  Our proposed
robustness measure is less conservative as compared to the existing
probabilistic robustness measure.  The fundamental connection
between our measure and the existing one is also established.

\appendix

\section{Proof of Theorem 1}

The following Lemma \ref{uniform_rad} is due to \cite {BLT}.  \beL
\la{uniform_rad} For any robustness requirement,
$\inf_{f_{\bs{\vDe}} \in \mscr{G}} \Pr \{ \bb{I} (\bs{\vDe}) = 1
\mid ||\bs{\vDe}|| \leq \ga  \} = \underline{\bb{P}} (\ga)$.
\eeL

\beL  \la{include} $\mscr{G}$ is a superset of $\mscr{F}$, i.e., $\mscr{G} \supseteq \mscr{F}$.  \eeL

\bpf Let $f_{\bs{\vDe}} \in \mscr{F}$.  We need to show
$f_{\bs{\vDe}} \in \mscr{G}$.  Let $0 < r_1 < r_2$ be two numbers
such that, for any $\vDe_1, \; \vDe_2$ satisfying $||\vDe_1|| = r_1,
\; ||\vDe_2|| = r_2$, both $f_{\bs{\vDe}} (\vDe_1)$ and
$f_{\bs{\vDe}} (\vDe_2)$ exist.  By the radial symmetry of the
distribution of $\bs{\vDe}$, we can write $f_{\bs{\vDe}} (\vDe_i)$
as $g(r_i)$ for $i =1, \; 2$. Clearly, the existence implies that
$g(.)$ is continuous at $r = r_i, \; i = 1 ,2$.   Let $c = \int_{v
\in \mcal{B}_1} d v$. By the radial symmetry of the distribution of
$\bs{\vDe}$ and the scaling property of the function $||.||$, we
have $f_{||\bs{\vDe}||} (r_i) = \lim_{\vep \to 0} \f{1}{2 \vep}
\int_{r_i - \vep}^{r_i + \vep} g(\ro)  \; n c \ro^{n-1} \; d \ro$
for $i = 1, 2$, where $n$ is the dimension of $\bs{\vDe}$.  Hence,
$f_{||\bs{\vDe}||} (r)$ is continuous at $r = r_i, \; i = 1 ,2$.
Recall that $f_{\bs{\vDe}} \in \mscr{F}$, we have $f_{||\bs{\vDe}||}
(r_1) \geq f_{||\bs{\vDe}||} (r_2)$.  On the other hand, by the
radial symmetry of the distribution of $\bs{\vDe}$ and the scaling
property of the function $||.||$, we have $g(r_i) = \lim_{\vep
\downarrow 0} \f{ \int_{r_i - \vep}^{r_i + \vep}  f_{||\bs{\vDe}||}
(\ro) \; d \ro }{c (r_i + \vep)^{n} -  c (r_i - \vep)^{n} }$ for $i
= 1, 2$.  By the continuity of $f_{||\bs{\vDe}||} (r)$ at $r_i$, we
have $g(r_i) = \f{ f_{||\bs{\vDe}||} (r_i) } { n \; c \; r_i^{n-1}}$
for $i = 1, 2$. It follows that $\f{ g(r_1) } { g(r_2) } = \li( \f{
r_2 } { r_1  } \ri )^{n-1} \f{ f_{||\bs{\vDe}||} (r_1) } {
f_{||\bs{\vDe}||} (r_2)  } \geq \li( \f{ r_2 } { r_1  } \ri )^{n-1}
\geq 1$, implying that $f_{\bs{\vDe}} \in \mscr{G}$. Hence,
$\mscr{G} \supseteq \mscr{F}$.

\epf

\beL \la{scale} For any $S \subseteq \pa \mcal{B}$, $\mrm{area}(S_r)
= r^{n-1} \; \mrm{area}(S)$ where $S_r = \{r \vDe: \; \vDe \in S\}$
and $n$ is the dimension of $\mcal{B}$.

\eeL

\bpf

By the scalable property of $||.||$, {\small \bee \li \{\f{\ro}{r}
\vDe :  r - \vep_1 < \ro < r + \vep_2, \; \vDe \in S_r \ri \} & = &
\li \{\ro \vDe :  r - \vep_1 < \ro < r + \vep_2, \; \vDe \in  S \ri
\} = \li \{\vDe :  r - \vep_1 < \ro < r + \vep_2, \; \f{\vDe}{\ro}
\in  S \ri \}. \eee} Hence, by invoking the definition
(\ref{defarea}), $\mrm{area}(S_r) = \lim_{\vep_1 \downarrow 0 \atop{
\vep_2 \downarrow 0 } } \f{ \int_{ q \in \li \{\vDe : \; r - \vep_1
< \ro < r + \vep_2, \; \f{\vDe}{\ro} \in S \ri \} } d q } { \vep_1 +
\vep_2 }$.  Making a change of variable $q = r q^\prime$ yields
 \bee \mrm{area}(S_r)
& = &  r^n \lim_{\vep_1 \downarrow 0 \atop{ \vep_2 \downarrow 0 } }
\f{ \int_{ q^\prime \in \li \{\vDe : \; r - \vep_1 < \ro < r +
\vep_2, \; \f{r \vDe}{\ro} \in S \ri \} } d
q^\prime } { \vep_1 + \vep_2 }\\
& = &  r^{n-1} \lim_{\vep_1 \downarrow 0 \atop{ \vep_2 \downarrow 0
} }
 \f{ \int_{ q^\prime \in \li \{\f{\ro}{r} \vDe : \; -
\f{\vep_1  }{r} \leq \f{\ro}{r} - 1 \leq
\f{\vep_2}{r}, \; \vDe \in S \ri \} } d q^\prime } { (\vep_1 + \vep_2) \sh r }\\
& = &  r^{n-1} \lim_{\vep_1 \downarrow 0 \atop{ \vep_2 \downarrow 0
} }
 \f{ \int_{ q^\prime \in \li \{\ro \vDe : \; - \vep_1
\leq \ro - 1  \leq \vep_2, \; \vDe \in S \ri \}
} d q^\prime } { \vep_1 + \vep_2 }\\
& = & r^{n-1} \mrm{area}(S). \eee

\epf

\beL \la{eqi} Suppose the distribution of $\bs{\vDe}$ is radially
symmetrical.  Let $S$ be a subset of $\pa \mcal{B} = \{ \vDe: ||\vDe
|| = 1 \}$. Then,  $\Pr \li \{  \li. \f{ \bs{\vDe} } { ||
\bs{\vDe}||} \in S \; \ri | \;  || \bs{\vDe}|| = \ro \ri \} = \f{
\mrm{area}(S) } {\mrm{area} (\pa \mcal{B}) }$ for any $\ro
> 0$ such that $f_{|| \bs{\vDe}||} (\ro)$ is
continuous. \eeL

 \bpf   By the definition of the conditional probability,
\[
\Pr \li \{  \li. \f{ \bs{\vDe} } { || \bs{\vDe}||} \in S \; \ri | \;
|| \bs{\vDe}|| = \ro \ri \} = \lim_{\vep_1 \downarrow 0 \atop{
\vep_2 \downarrow 0 } }
 \f{ \Pr \li \{ \f{ \bs{\vDe} } { ||
\bs{\vDe}||} \in S,  \;  \ro - \vep_1 \leq || \bs{\vDe}|| \leq \ro +
\vep_2 \ri \} } {\Pr \li \{ \ro - \vep_1 \leq || \bs{\vDe}|| \leq
\ro + \vep_2 \ri \} }.
\]
We claim that $\li \{ \f{ \bs{\vDe} } { || \bs{\vDe}||} \in S,  \;
\ro - \vep_1 \leq || \bs{\vDe}|| \leq \ro + \vep_2 \ri \} = \{
\bs{\vDe} \in S_{\ro, \vep_1, \vep_2} \}$ where $S_{\ro, \vep_1,
\vep_2} = \{ \vDe : \f{\vDe}{\ro^\prime} \in S, \; \ro - \vep_1 \leq
\ro^\prime \leq \ro + \vep_2 \}$.  To show this claim, it suffices
to show that $\li \{ \vDe : \f{ \vDe } { || \vDe ||} \in S, \; \ro -
\vep_1 \leq || \vDe || \leq \ro + \vep_2 \ri \} =  S_{\ro, \vep_1,
\vep_2}$.   Let $\vDe \in S_{\ro, \vep_1, \vep_2}$. By definition,
there exists $\ro^\prime \in [\ro - \vep_1, \ro + \vep_2 ]$ such
that $\f{\vDe}{\ro^\prime} \in S$. Therefore, by the scalable
property of the function $||.||$, we have $|| \vDe|| = \li | \li
|\ro^\prime  \f{\vDe}{\ro^\prime} \ri | \ri | = \ro^\prime \li | \li
| \f{\vDe}{\ro^\prime} \ri | \ri | = \ro^\prime \in [\ro - \vep_1,
\ro + \vep_2]$ and $\f{ \vDe } { || \vDe||} = \f{
\f{\vDe}{\ro^\prime} } { \f{||\vDe||}{\ro^\prime}}
 =\f{ \f{\vDe}{\ro^\prime} } { \li | \li | \f{\vDe}{\ro^\prime} \ri | \ri |}
= \f{\vDe}{\ro^\prime} \in S$.  This implies that $\vDe \in  \li \{
\vDe : \f{ \vDe } { || \vDe ||} \in S, \; \ro - \vep_1 \leq || \vDe
|| \leq \ro + \vep_2 \ri \}$.

Now let $\vDe \in \li \{ \vDe : \f{ \vDe } { || \vDe ||} \in S, \;
\ro - \vep_1 \leq || \vDe || \leq \ro + \vep_2 \ri \}$ and
$\ro^\prime = ||\vDe||$. By definition, $\ro - \vep_1 \leq
\ro^\prime \leq \ro + \vep_2, \; \f{\vDe}{\ro^\prime} \in S$. Hence,
$\vDe \in S_{\ro, \vep_1, \vep_2}$.  The claim is thus proved and we
have
\[
\Pr \li \{ \f{ \bs{\vDe} } { || \bs{\vDe}||} \in S,  \;  \ro -
\vep_1 \leq || \bs{\vDe}|| \leq \ro + \vep_2 \ri \} = \Pr \{
\bs{\vDe} \in S_{\ro, \vep_1, \vep_2} \}.
\]
Let $S_{\ro^\prime} = \{\ro^\prime \vDe : \vDe \in S \}$. Then,
$S_{\ro^\prime}\subseteq \pa \mcal{B}_{\ro^\prime}$ and $S_{\ro,
\vep_1, \vep_2} = \{\vDe: \vDe \in S_{\ro^\prime}, \ro - \vep_1 \leq
\ro^\prime \leq \ro + \vep_2 \}$.  By the notion of the radially
symmetrical distribution of $\bs{\vDe}$ and the property of the area
function shown in Lemma \ref{scale}, we have $\Pr \{ \bs{\vDe} \in
S_{\ro^\prime} \mid ||\bs{\vDe}|| = \ro ^\prime  \}  =  \f{
\mrm{area}(S_{\ro^\prime}) } {\mrm{area} (\pa \mcal{B}_{\ro^\prime})
} = \f{ {\ro^\prime}^{n-1} \mrm{area}(S) } { {\ro^\prime}^{n-1}
\mrm{area} (\pa \mcal{B}) } = \f{ \mrm{area}(S) } {\mrm{area} (\pa
\mcal{B}) }$. On the other hand, by the definition of the
conditional probability,
\[
\Pr \{ \bs{\vDe} \in S_{\ro^\prime} \mid ||\bs{\vDe}|| = \ro
^\prime  \} = \lim_{\vep_1 \downarrow 0 \atop{ \vep_2 \downarrow 0
} } \f{ \Pr \{ \bs{\vDe} \in S_{\ro, \vep_1, \vep_2} \}  } {\Pr
\li \{ \ro - \vep_1 \leq || \bs{\vDe}|| \leq \ro + \vep_2 \ri \}
}.
\]
It follows that $\Pr \li \{  \li. \f{ \bs{\vDe} } { || \bs{\vDe}||}
\in S \; \ri | \;  || \bs{\vDe}|| = \ro \ri \} =  \lim_{\vep_1
\downarrow 0 \atop{ \vep_2 \downarrow 0 } } \f{ \Pr \{ \bs{\vDe} \in
S_{\ro, \vep_1, \vep_2} \}  } {\Pr \li \{ \ro - \vep_1 \leq ||
\bs{\vDe}|| \leq \ro + \vep_2 \ri \} } = \f{ \mrm{area}(S) }
{\mrm{area} (\pa \mcal{B}) }$.

\epf

\beL \la{lem55} Suppose $f_{||\bs{\vDe}||}(.)$ is continuous in $(a,
b)$. Then,
 $\Pr \li \{ \f{ \bs{\vDe} } { || \bs{\vDe}||} \in S \mid a < ||
\bs{\vDe}|| < b \ri \} = \f{ \mrm{area}(S) } {\mrm{area} (\pa
\mcal{B}) }$.  \eeL

\bpf

Let $\eta > 0$ and $\de \in \li ( 0, \f{b-a}{2} \ri )$.  For
notational simplicity, let $c = \f{ \mrm{area}(S) } {\mrm{area} (\pa
\mcal{B}) }$.  By Lemma \ref{eqi}, for any $\ro \in [a + \de, b -
\de]$, we can find $\vep = \vep(\ro)$ such that $\li | \Pr \li \{
\f{ \bs{\vDe} } { || \bs{\vDe}||} \in S \mid \ro - \vep_1 \leq ||
\bs{\vDe}|| \leq \ro + \vep_2 \ri \}  - c \ri | < \eta$ for any
positive $\vep_1, \vep_2$ less than $\vep(\ro)$.  Hence, the union
of the open intervals $\cup_{\ro \in [a+ \de, b- \de]} (\ro -
\vep(\ro), \ro + \vep(\ro))$ will cover interval $[a+ \de, b -
\de]$.  By the {\it finite coverage theorem}, we can choose finite
number of $\ro_i$ from $[a+ \de, b - \de]$ such that $\cup_{i = 1}^k
(\ro_i - \vep(\ro_i), \ro_i + \vep(\ro_i))$ covers interval $[a +
\de, b - \de]$ and that none of $(\ro_i - \vep(\ro_i), \ro_i +
\vep(\ro_i))$ is nested in another. By using the mid-points of the
intersections of every two consecutive intervals as dividing points,
we can partition $[a + \de, b - \de]$ as $k$ intervals $[a_i, b_i]$
such that $\li | \Pr \li \{ \f{ \bs{\vDe} } { || \bs{\vDe}||} \in S
\mid a_i \leq || \bs{\vDe}|| \leq b_i \ri \} - c \ri | < \eta$ for
$i = 1, \cd, k$.  Therefore, $\li | \Pr \li \{ \f{ \bs{\vDe} } { ||
\bs{\vDe}||} \in S,  \;  a_i \leq || \bs{\vDe}|| \leq b_i \ri \} - c
\Pr \li \{ a_i \leq || \bs{\vDe}|| \leq b_i \ri \} \ri | < \eta \Pr
\li \{ a_i \leq || \bs{\vDe}|| \leq b_i \ri \}$ for $i = 1, \cd, k$
and
\[
\li | \sum_{i=1}^k \li [ \Pr \li \{ \f{ \bs{\vDe} } { || \bs{\vDe}||} \in S,  \;  a_i \leq || \bs{\vDe}|| \leq
b_i \ri \} - c \Pr \li \{ a_i \leq || \bs{\vDe}|| \leq b_i \ri \} \ri ] \ri | < \eta \sum_{i=1}^k \Pr \li \{ a_i
\leq || \bs{\vDe}|| \leq b_i \ri \}.
\]
That is, $\li | \Pr \li \{ \f{ \bs{\vDe} } { || \bs{\vDe}||} \in S,
\;  a + \de \leq || \bs{\vDe}|| \leq b - \de \ri \} - c \Pr \li \{ a
+ \de \leq || \bs{\vDe}|| \leq b - \de \ri \} \ri | < \eta \Pr \li
\{ a + \de  \leq || \bs{\vDe}|| \leq b - \de \ri \}$. As a result,
$\li | \Pr \li \{ \f{ \bs{\vDe} } { || \bs{\vDe}||} \in S \mid a +
\de \leq || \bs{\vDe}|| \leq b - \de \ri \} - c \ri | < \eta$. Since
$\eta$ can be arbitrarily small, we have
\[
\Pr \li \{ \f{ \bs{\vDe} } { || \bs{\vDe}||} \in S,  \; a + \de
\leq || \bs{\vDe}|| \leq b - \de \ri \} = c \Pr \li \{ a + \de
\leq || \bs{\vDe}|| \leq b - \de \ri \}.
\]
By the assumption that $f_{||\bs{\vDe}||}(.)$ is piece-wise
continuous, we have $\Pr \li \{ \ro \leq || \bs{\vDe}|| \leq \ro +
\de \ri \} \to 0$ as $\de \downarrow 0$ for all $\ro \geq 0$. Hence,
\bee &  & \lim_{\de \downarrow 0}  \li | \Pr \li \{ \f{ \bs{\vDe} }
{ || \bs{\vDe}||} \in S,  \; a + \de \leq || \bs{\vDe}|| \leq b -
\de \ri \} - \Pr \li \{ \f{ \bs{\vDe} } { || \bs{\vDe}||} \in S,
\; a < || \bs{\vDe}|| <  b \ri \} \ri |\\
& = & \lim_{\de \downarrow 0} \li [ \Pr \li \{ \f{ \bs{\vDe} } { ||
\bs{\vDe}||} \in S,  \;  a < || \bs{\vDe}|| < a + \de \ri \} + \Pr
\li \{ \f{ \bs{\vDe} } { || \bs{\vDe}||} \in S,  \;  b - \de < ||
\bs{\vDe}|| < b \ri \} \ri ]\\
& \leq &  \lim_{\de \downarrow 0} \li [ \Pr \li \{ a < ||
\bs{\vDe}|| < a + \de \ri \} + \Pr \li \{ b - \de < || \bs{\vDe}|| <
b \ri \} \ri ]  = 0, \eee and so $\lim_{\de \downarrow 0}  \Pr \li
\{ \f{ \bs{\vDe} } { || \bs{\vDe}||} \in S,  \; a + \de \leq ||
\bs{\vDe}|| \leq b - \de \ri \} = \Pr \li \{ \f{ \bs{\vDe} } { ||
\bs{\vDe}||} \in S, \; a < || \bs{\vDe}|| <  b \ri \}$.  Similarly,
\bee &  & \lim_{\de \downarrow 0}  \li | \Pr \li \{  a + \de
\leq || \bs{\vDe}|| \leq b - \de \ri \} - \Pr \li \{ a < || \bs{\vDe}|| <  b \ri \} \ri |\\
& = & \lim_{\de \downarrow 0} \li [ \Pr \li \{ a < || \bs{\vDe}|| <
a + \de \ri \} + \Pr \li \{  b - \de < || \bs{\vDe}|| < b \ri \} \ri
] = 0, \eee and so $\lim_{\de \downarrow 0}  \Pr \li \{ a + \de \leq
|| \bs{\vDe}|| \leq b - \de \ri \} = \Pr \li \{ a < || \bs{\vDe}|| <
b \ri \}$.  It follows that
\[
\Pr \li \{ \f{ \bs{\vDe} } { || \bs{\vDe}||} \in S,  \; a < ||
\bs{\vDe}|| < b \ri \} = c \Pr \li \{ a < || \bs{\vDe}|| < b \ri
\}.
\]
This completes the proof.

 \epf

\beL \la{independence} Suppose that the distribution of $\bs{\vDe}$
is radially symmetrical and that $f_{||\bs{\vDe}||}(.)$ is
piece-wise continuous over $(0, \iy)$.  Then, $\f{ \bs{\vDe} } { ||
\bs{\vDe}||}$ is independent with $|| \bs{\vDe}||$.  Moreover, $\f{
\bs{\vDe} } { || \bs{\vDe}||}$ is uniformly distributed over $\{
\vDe: ||\vDe || = 1 \}$.  \eeL

\bpf

 Since $f_{||\bs{\vDe}||}(.)$ is piece-wise continuous over
$(0, \iy)$, we can represent $(0, \iy)$ as a union of open intervals
$(a_i, b_i)$ where $f_{||\bs{\vDe}||}(.)$ is continuous and the set
of discrete values $\ro_j, \; j = 1, 2, \cd$ for which
$f_{||\bs{\vDe}||}(.)$ is discontinuous. We can enumerate the
intervals and the discrete values such that $b_i - a_i$ is
non-increasing with respect to $i$ and that $\ro_j - \ro_{j-1}$ is
non-increasing with respect to $j$. Then, $\Pr \li \{ \f{ \bs{\vDe}
} { || \bs{\vDe}||} \in S, \; || \bs{\vDe}|| = \ro_j \ri \} = 0, \;
j = 1, 2, \cd$ and, by Lemma \ref{lem55},

\bee \Pr \li \{ \f{ \bs{\vDe} } { || \bs{\vDe}||} \in S \ri \} & =
& \sum_{i} \Pr \li \{ \f{ \bs{\vDe} } { || \bs{\vDe}||} \in S, \;
a_i < || \bs{\vDe}|| < b_i \ri \} + \sum_j \Pr \li \{ \f{
\bs{\vDe} } { || \bs{\vDe}||} \in S, \; || \bs{\vDe}|| = \ro_j \ri
\}\\
& = & \f{ \mrm{area}(S) } {\mrm{area} (\pa \mcal{B}) } \li [
\sum_{i} \Pr \li \{ a_i < || \bs{\vDe}|| < b_i \ri \} +  \sum_j \Pr
\li \{ || \bs{\vDe}|| = \ro_j \ri \} \ri ] = \f{ \mrm{area}(S) }
{\mrm{area} (\pa \mcal{B}) }. \eee

Therefore, invoking Lemma \ref{eqi}, we have $\Pr \li \{ \f{
\bs{\vDe} } { || \bs{\vDe}||} \in S \mid || \bs{\vDe}|| = \ro \ri \}
= \Pr \li \{ \f{ \bs{\vDe} } { || \bs{\vDe}||} \in S \ri \}$ for any
$\ro$ such that $f_{||\bs{\vDe}||}(.)$ is continuous.  This implies
the independence between $\f{ \bs{\vDe} } { || \bs{\vDe}||}$ and $||
\bs{\vDe}||$. Moreover, since the argument holds for any $S
\subseteq \{ \vDe : ||\vDe|| = 1 \}$, we have that $\f{ \bs{\vDe} }
{ || \bs{\vDe}||}$ is uniformly distributed over $\{ \vDe : ||\vDe||
= 1 \}$.  The proof is thus completed.

 \epf

 \beL \la{eqphi} Suppose that $\phi(.)$ is continuous over $(a, b)$ and that the distribution of uncertainty $\bs{\vDe}$ is radially
 symmetrical and continuous over $(a, b)$. Then $\Pr \{ \bb{I}(\bs{\vDe} ) = 1, \; a < || \bs{\vDe} || < b \} =
\int_a^b \phi(r) f_R(r) d r$.
 \eeL

 \bpf
Define $U = \f{ \bs{\vDe} } { || \bs{\vDe}||}, \; R = ||\bs{\vDe}||$
and $f_R(\ro) = \f{d \li [ \Pr \{ R \leq \ro  \}  \ri ] } { d \ro
}$. By Lemma \ref{independence}, we have that $U$ and $R$ are
independent and that $U$  is uniform over $\pa \mcal{B}$. Hence, the
probability density function of $UR$ is $\f{1}{ \mrm{area}(\pa
\mcal{B}) } \times f_R(r)$ and, by the Fubini's Theorem, \bee \Pr \{
\bb{I}(\bs{\vDe} ) = 1, \; a <
|| \bs{\vDe} || < b \} & = & \Pr \{ \bb{I}(U R) = 1, \; a  < R < b \}\\
& = & \int_{r = a}^b \int_{\{ u : \; \bb{I}(r u) = 1, \; u \in \pa
\mcal{B}
\} } \f{1}{ \mrm{area}(\pa \mcal{B}) } f_R(r) \; d u d r\\
 & = & \int_{r = a}^b \li [ \int_{\{ u : \; \bb{I}(r u) = 1, \; u \in \pa \mcal{B}
\} } \f{1}{ \mrm{area}(\pa \mcal{B}) } d u  \ri ] f_R(r) d r\\
& = &  \int_{r = a}^b \phi(r) f_R(r) d r \eee where the last
equality follows from the definition of $\phi(.)$.

\epf

\beL \la{lemdec} Suppose that $\phi(.)$ is piece-wise continuous and
that $f_{|| \bs{\vDe} ||}(.)$ is piece-wise continuous and
non-increasing. Then, $\Pr \{  \bb{I} (\bs{\vDe}) = 1, \;
||\bs{\vDe}|| \leq \ga \} = \int_{  0}^\ga \phi(\ro) \; f_{||
\bs{\vDe} ||}(\ro) d \ro$. \eeL

\bpf

Let $\vep > 0$.  Since $f_{|| \bs{\vDe} ||}(\ro)$ is non-increasing,
we have $f_{|| \bs{\vDe} ||}(\ro) \leq \f{ \Pr \{ || \bs{\vDe} ||
\leq \vep \} } { \vep }$ for $\ro \in [\vep, \ro]$. It follows that
$\phi(\ro) \; f_{|| \bs{\vDe} ||}(\ro)$ is piece-wise continuous and
bounded for $\ro \in [\vep, \ro]$. Hence, the Riemann integral
$\int_{ \vep}^\ga \phi(\ro) \; f_{|| \bs{\vDe} ||}(\ro) d \ro$
exists.  Note that $\int_{  \vep}^\ga \phi(\ro) \; f_{|| \bs{\vDe}
||}(\ro) d \ro \leq \int_{  \vep}^\ga  f_{|| \bs{\vDe} ||}(\ro) d
\ro \leq 1$ and that $\int_{  \vep}^\ga \phi(\ro) \; f_{|| \bs{\vDe}
||}(\ro) d \ro$ is non-increasing with respect to $\vep$. Thus,
$\lim_{\vep \downarrow 0} \int_{  \vep}^\ga \phi(\ro) \; f_{||
\bs{\vDe} ||}(\ro) d \ro$ exists.  This limit is denoted as $\int_{
0}^\ga \phi(\ro) \; f_{|| \bs{\vDe} ||}(\ro) d \ro$.

Note that we can partition interval $(0, \ga)$ as a sequence of
intervals $(a_i, b_i), \; i = 1, \cd, \iy$ such that $a_i, \; b_i,
\; i = 1, 2, \cd$ are discontinuities of $f_{|| \bs{\vDe} ||}(\ro)$
and that $b_i - a_i$ is non-increasing with respect to $i$. To
ensure that the partition is unique, we can handle the situation
that some intervals have the same length by enforcing the following
criterion: if $b_i - a_i = b_j - a_j, \; i < j$ then $a_i < a_j$.
Then, by the property of the Riemann integral, we have $\int_{
0}^\ga \phi(\ro) \; f_{|| \bs{\vDe} ||}(\ro) d \ro = \sum_{i=1}^\iy
\int_{ a_i}^{b_i} \phi(\ro) \; f_{|| \bs{\vDe} ||}(\ro) d \ro$. On
the other hand, since $\Pr \{ ||\bs{\vDe}|| = a_i \} = \Pr \{
||\bs{\vDe}|| = b_i \} = 0$ for $i = 1, 2 \cd, \iy$, we have \bel
\Pr \{  \bb{I} (\bs{\vDe}) = 1, \; ||\bs{\vDe}|| \leq \ga \} & = &
\sum_{i=1}^\iy \Pr \{  \bb{I}
(\bs{\vDe}) = 1, \;  a_i < ||\bs{\vDe}|| < b_i \} \nonumber\\
& = & \sum_{i=1}^\iy \int_{  a_i}^{b_i} \phi(\ro) \; f_{||
\bs{\vDe} ||}(\ro) d \ro \la{use5} \\
& = & \int_{  0}^{\ga} \phi(\ro) \; f_{|| \bs{\vDe} ||}(\ro) d \ro
\nonumber
 \eel
 where the equality (\ref{use5}) follows from Lemma \ref{eqphi}.

\epf

\beL \la{chen} For any $r > 0$, $\mscr{P}(r) = \f{1}{r} \int_{  0}^r
\phi(\ro)  \; d \ro$. \eeL

\bpf

By the definition of $\mscr{P}(.)$, we have $\mscr{P}(r) = \Pr \{
\bb{I}(U R) = 1 \} = \Pr \{ \bb{I}(U R) = 1, \; ||UR|| \leq r \}$
where $U$ and $R$ are independent random variables such that $U$ is
uniformly distributed over $\pa \mcal{B}$ and $R$ is uniformly
distributed over $[0, r]$.  Applying Lemma \ref{lemdec} to random
variable $\bs{\vDe} = U R$, we have $\mscr{P}(r) = \int_{  0}^r
\phi(\ro) \; f_R(\ro) \; d \ro = \f{1}{r} \int_{  0}^r \phi(\ro) \;
d \ro$.

\epf

\beL \la{debound} Let $0 < r_1 < r_2$.  Then, $| \mscr{P}(r_2) -
\mscr{P}(r_1) | < \f{ 2 (r_2 - r_1) }{ r_1 }$. \eeL

\bpf By Lemma \ref{chen},  \bee | \mscr{P}(r_2) - \mscr{P}(r_1) | &
= & \li | \f{ \int_{r_1}^{r_2} \phi(\ro) \; d \ro } {r_2 } + \li
(\f{1}{r_2} -
\f{1}{r_1}\ri ) \int_0^{r_1} \phi(\ro) \; d \ro  \ri |\\
& \leq &  \f{ \int_{r_1}^{r_2} \phi(\ro) \; d \ro } {r_2 }  + \li
(\f{1}{r_1} - \f{1}{r_2}\ri ) \int_0^{r_1} \phi(\ro) \; d \ro\\
& \leq & \f{ r_2 - r_1 } { r_2 } +   \f{ r_2 - r_1 } { r_1 r_2 }
r_1\\
& = & \f{ 2 (r_2 - r_1) }{ r_2 } \leq  \f{ 2 (r_2 - r_1) }{ r_1 }
\eee where we have used the fact that $0 \leq \phi(\ro) \leq 1$.

\epf

\beL \la{dense} $\inf_{0 < \ro \leq \ga \atop{ \f{\ro}{\ga} \in
\bb{Q}} } \mscr{P}(\ro) = \inf_{0 < \ro \leq \ga} \mscr{P}(\ro)$
where $\bb{Q}$ denotes the set of all rational numbers.

\eeL

\bpf Let $a = \inf_{0 < \ro \leq \ga \atop{ \f{\ro}{\ga} \in \bb{Q}}
} \mscr{P}(\ro)$ and $b = \inf_{0 < \ro \leq \ga} \mscr{P}(\ro)$.
Clearly, $a \geq b \geq 0$. Suppose $a > b$. Then, there exists a
real  number $\ro^* \in (0, \ga]$ such that $\mscr{P}(\ro^*) < \f{a
+ b}{2}$.  By the dense property of the rational numbers, for any
$\de \in (0, \ro^*)$, there exists a number $\se$ such that
$\f{\se}{\ga} \in \bb{Q}$ and that $\li | \se - \ro^* \ri | < \de$.
Thus, by Lemma \ref{debound}, $| \mscr{P}(\se) - \mscr{P}(\ro^*)|
\leq \f{2 \de}{\ro^* - \de}$, leading to $\mscr{P}(\se) \leq
\mscr{P}(\ro^*) + \f{2 \de}{\ro^* - \de} < \f{a + b}{2} + \f{2
\de}{\ro^* - \de}$.  Since $\de$ can be arbitrarily small, we have
$\mscr{P}(\se) \leq \f{a + b}{2}$.  Hence, $a \leq \f{a + b}{2}$,
i.e., $a \leq b$, contradicting to $a > b$.   This shows that $a >
b$ is not true.  Therefore, $a = b$.

 \epf

We are now in the position to prove Theorem 1.  For every $f_{
\bs{\vDe} } \in \mscr{F}$, define {\small $f_{|| \bs{\vDe} ||}(\ro,
\ga) = \f{d \; \Pr \{ || \bs{\vDe} || \leq \ro \; \mid \; ||
\bs{\vDe} || \leq \ga \} } {d \ro }$}.  Then, $f_{|| \bs{\vDe}
||}(\ro, \ga) = \f{1}{\Pr \{ || \bs{\vDe} || \leq \ga \}} \f{d \;
\Pr \{ || \bs{\vDe} || \leq \ro \} } {d \ro } = \f{ f_{|| \bs{\vDe}
||}(\ro) } { \Pr \{ || \bs{\vDe} || \leq \ga \} }$, and the set of
all such functions constitute a family of conditional density
functions, denoted by $\mscr{F}_\ga$.  Clearly, every conditional
density $f_{|| \bs{\vDe} ||}(\ro, \ga)$ in $\mscr{F}_\ga$ is
non-increasing with respect to $\ro$.  For every positive integer
$k$, we use $\mscr{F}_{\ga, k}$ to denote the set of conditional
density functions of the form: $f_{|| \bs{\vDe} ||}(\ro, \ga) =
\sum_{i=1}^k \xi_i \; I_{(r_{i-1}, r_i]} (\ro), \; \fa \ro \in (0,
\ga]$ where $r_i = \frac{i \; \ga}{k}, \; i = 0, 1 , \cdots, k$,
\[
I_{(r_{i-1}, r_i]} (x) = \bec 1 & \tx{if} \; x \in
(r_{i-1}, r_i];\\
0 & \tx{otherwise} \eec
\]
and $\xi_1 \geq \xi_2 \geq \cdots \geq \xi_k  \geq 0$ with
$\frac{\ga}{k} \sum_{i=1}^k \xi_i = 1$.  By Lemma \ref{lemdec},
 \bee \Pr \{  \bb{I} (\bs{\vDe}) = 1 \mid ||\bs{\vDe}|| \leq \ga \} & =
 & \f{\Pr \{  \bb{I} (\bs{\vDe}) = 1, \; ||\bs{\vDe}|| \leq \ga
 \}}{ \Pr \{ || \bs{\vDe} || \leq \ga \} }  =  \f{\int_{ 0}^\ga \phi(\ro) \; f_{|| \bs{\vDe} ||}(\ro)
 d \ro}{ \Pr \{ || \bs{\vDe} || \leq \ga \} } =  \int_{  0}^\ga  \phi(\ro) f_{|| \bs{\vDe} ||}(\ro, \ga)
 d \ro.
\eee Therefore, \be \la{bri}
\inf_{ f_{\bs{\vDe}} \in \mscr{F} } \Pr
\{ \bb{I} (\bs{\vDe}) = 1 \mid ||\bs{\vDe}|| \leq \ga \} = \inf_{
f_{|| \bs{\vDe} ||} (., \ga) \in \mscr{F}_\ga } \int_{0}^\ga
\phi(\ro) f_{|| \bs{\vDe} ||}(\ro, \ga) d \ro. \ee Since $\phi(\ro)
 \;  I_{(r_{i-1},
r_i]} (\ro)$ is bounded and piece-wise continuous over $(0,\ga]$, it
is Riemann integrable.  It follows that, for a conditional density
$f_{|| \bs{\vDe} ||}(\ro, \ga)$ in the family $\mscr{F}_{\ga, k}$,
 \bee \int_{0}^\ga \phi(\ro) f_{|| \bs{\vDe} ||}(\ro, \ga) d \ro & = &
\int_{  0}^\ga \phi(\ro) \; \li [ \sum_{i=1}^k \xi_i \; I_{(r_{i-1},
r_i]} (\ro) \ri ]
 d \ro  =  \sum_{i=1}^k  \li [ \int_{  0}^\ga  \phi(\ro)
 \;  I_{(r_{i-1},
r_i]} (\ro) \; d \ro \ri ] \xi_i   =  \sum_{i=1}^k a_i \; \xi_i \eee
where $a_i = \int_{  0}^\ga \phi(\ro)
 \;  I_{(r_{i-1},
r_i]} (\ro) \; d \ro$ for $i = 1, \cd, k$.  Since $a_i$ is
independent of $(\xi_1, \cd, \xi_k)$ for $i = 1, \cd, k$, we have
that $\sum_{i=1}^k a_i \; \xi_i$ is a linear function of $\xi_i, \;
i = 1, \cd, k$ for any given $k > 0$.  Therefore, the infimum
$\inf_{ f_{|| \bs{\vDe} ||} (., \ga) \in \mscr{F}_{\ga, k} }
\int_{0}^\ga \phi(\ro) f_{|| \bs{\vDe} ||}(\ro, \ga) d \ro$ equals
to the minimum of $\sum_{i=1}^k a_i \; \xi_i$ subject to the
constraint that $\xi_1 \geq \xi_2 \geq \cdots \geq \xi_k  \geq 0$
and $\frac{\ga}{k} \sum_{i=1}^k \xi_i = 1$. Note that the minimum of
a linear program over a bounded set is achieved at the extreme
points. By Lemma 2.2 of \cite{BL}, for every extreme point of the
convex set $\{ (\xi_1, \cd, \xi_k): \xi_1 \geq \xi_2 \geq \cdots
\geq \xi_k \geq 0, \; \frac{\ga}{k} \sum_{i=1}^k \xi_i = 1 \}$, we
can find an integer $\ell$ such that $\xi_i = \f{k } { \ga \ell }$
for $i = 1, \cd, \ell$ and $\xi_i = 0$ for $i = \ell + 1, \cd, k$.
For such extreme point associated with $\ell$, we have $\sum_{i=1}^k
a_i \; \xi_i = \int_{0}^\ga \phi(\ro) f_{|| \bs{\vDe} ||}(\ro, \ga)
d \ro = \int_{0}^{\f{\ell}{k} \ga} \phi(\ro) \f{ 1 } { \f{\ell}{k}
\ga } d \ro =  \mscr{P} \li ( \f{\ell}{k} \ga \ri )$, where the last
equality follows from Lemma \ref{chen}.  Therefore,
\[
\inf_{ f_{|| \bs{\vDe} ||} (., \ga) \in \mscr{F}_{\ga, k} }
\int_{0}^\ga \phi(\ro) f_{|| \bs{\vDe} ||}(\ro, \ga) d \ro = \min
\li \{ \mscr{P} \li ( \f{\ell}{k} \ga \ri ): 0 \leq \ell \leq k \ri
\}.
\]
It follows that {\small \[ \inf_{ f_{|| \bs{\vDe} ||} (., \ga) \in
\cup_{k=1}^\iy \mscr{F}_{\ga, k} } \int_{0}^\ga \phi(\ro) f_{||
\bs{\vDe} ||}(\ro, \ga) d \ro = \inf \bigcup_{k=1}^\iy \li \{
\mscr{P} \li ( \f{\ell}{k} \ga \ri ): 0 \leq \ell \leq k \ri \} =
\inf \li \{ \mscr{P}(\ro) : 0 < \ro \leq \ga, \; \f{\ro}{\ga} \in
\bb{Q} \ri \}.
\]}
It can be shown that
\[
\inf_{ f_{|| \bs{\vDe} ||} (., \ga) \in \cup_{k=1}^\iy
\mscr{F}_{\ga, k} } \int_{0}^\ga \phi(\ro) f_{|| \bs{\vDe} ||}(\ro,
\ga) d \ro = \inf_{ f_{|| \bs{\vDe} ||} (., \ga) \in \mscr{F}_{\ga}
} \int_{0}^\ga \phi(\ro) f_{|| \bs{\vDe} ||}(\ro, \ga) d \ro.
\]
Hence, by (\ref{bri}),
\[
\inf_{ f_{\bs{\vDe}} \in \mscr{F} } \Pr \{ \bb{I} (\bs{\vDe}) = 1
\mid ||\bs{\vDe}|| \leq \ga \} = \inf \li \{ \mscr{P}(\ro) : 0 < \ro
\leq \ga, \; \f{\ro}{\ga} \in \bb{Q} \ri \} = \inf_{0 < \ro \leq
\ga} \mscr{P}(\ro),
\]
where the last equality follows from Lemma \ref{dense}.  Finally, by
Lemma \ref{uniform_rad} and Lemma \ref{include}, we have
$\underline{\mscr{P}} (\ga) = \inf_{f_{\bs{\vDe}} \in \mscr{F}} \Pr
\{ \bb{I} (\bs{\vDe}) = 1  \mid ||\bs{\vDe}|| \leq \ga  \}  \geq
\inf_{f_{\bs{\vDe}} \in \mscr{G}} \Pr \{ \bb{I} (\bs{\vDe}) = 1 \mid
||\bs{\vDe}|| \leq \ga  \} =  \underline{\bb{P}} (\ga)$.   The proof
is thus completed.

\section{Proof of Theorem 2}

We shall first define some terminologies that will be used in the proof.

\begin{definition}
A value of the uncertainty radius is said to be a discontinuity if $\phi(.)$ is discontinuous for that value.
\end{definition}

\begin{definition}
An open interval $(a, b)$ is said to be a continuous interval if $\phi(r)$ is continuous for any $r \in (a, b)$.
\end{definition}

\begin{definition}
A discontinuity, $p$, is said to be a cluster point if, for any $\ep >0$, there exists another discontinuity,
$q$, such that $|p - q| < \ep$.
\end{definition}

The proof of the transform formulas is largely focused on the investigation of discontinuities, cluster points
and continuous intervals.  By the assumption that $\phi(.)$ is piece-wise continuous, we can see that the
distributions of discontinuities and cluster points can be arbitrary. For example, it is possible that there are
infinitely many discontinuities distributed over $(0, r)$ as $\f{r}{(i+1)(j+1) }$ where $i = 1, \cd, \iy$ and $j
= 1, \cd, \iy$. In this example, there are infinitely many cluster points $\f{r}{i+ 1 }, \; i = 1, \cd, \iy$.

Despite the complexity of the distributions of discontinuities and cluster points, it suffices to prove the
transform formulas for the following four cases:

\bed

\item [Case (1)] There are a finite number of discontinuities.

\item [Case (2)] There are infinitely many discontinuities such that $r= 0$ is the unique cluster point.

\item [Case (3)] There are infinitely many discontinuities such that there is a cluster point at $r= 0$
and that there is at least one more cluster point at $r > 0$.

\item [Case (4)] There are infinitely many discontinuities such that there is no cluster point at $r= 0$.

\eed

Before addressing each case in details, we need to establish some preliminary results.

The following lemma is on the enumeration and classification of
continuous intervals.

\beL \la{lemm4} For any $\vep > 0$, the set of all continuous
intervals defined by the end points $q, r$ or discontinuities of
interval $(q, r)$ can be divided into two classes such that i) the
first class, denoted by ${\wh{\mscr{I}}}_\vep$,  has a finite number
of intervals; ii) the second class, denoted by $\mscr{I}_\vep$, has
infinitely many intervals and the total length is less than $\vep$.
\eeL

\bpf Such classification can be performed as follows. Let $k = 1$
and $c_k = \f{1}{2^k}$.  Find all intervals with length greater than
$c_k$. Rank these intervals by the lengths and include it in set
$\mscr{A}$. Include the remaining intervals in set $\mscr{B}$.
Increment $k$ and update $c_k = \f{1}{2^k}$.  From $\mscr{B}$ find
all intervals with length greater than $c_k$. Add these intervals to
set $\mscr{A}$ and rank all intervals by the lengths. Eliminate
those intervals from set $\mscr{B}$.

Repeating these steps for infinitely many values of $k$ leads to a
sequence of intervals of decreasing lengths. Let $(a_i, \; b_i), \;
i = 1, 2, \cd$ denote this sequence.  Let $L_i = b_i - a_i$. Then,
$\sum_{i=1}^\iy L_i = r-q $ and $L_i$ is decreasing with respect to
$i$. Thus, by Cauchy's theorem, there must be an integer $K$ such
that $\sum_{i=K}^\iy L_i  < \vep$.  This implies that we have the
desired two classes.  The first class ${\wh{\mscr{I}}}_\vep$
consists of intervals $(a_i, b_i), \; i = 1, \cd, K-1$ and the
second class $\mscr{I}_\vep$ consists of intervals $(a_i, b_i), \; i
= K, \cd, \iy$. \epf

\beL \la{Barm} For any $r > 0$, $\bb{P}(r) = \f{n}{r^n} \int_{  0}^r
\phi(\ro) \;  \ro^{n-1} \; d \ro$ where $n$ is the dimension of
uncertainty space. \eeL

\bpf Since $\bs{\vDe}^{\mrm{u}}$ is uniformly distributed over
$\mcal{B}$, we can derive the density function of $||
\bs{\vDe}^{\mrm{u}} ||$ as $f_{ || \bs{\vDe}^{\mrm{u}} || } (\ro) =
\f{n \ro^{n-1}}{r^n}$.  By definition, $\bb{P}(r) = \Pr \{
\bb{I}(\bs{\vDe}^{\mrm{u}} ) = 1 \} = \Pr \{
\bb{I}(\bs{\vDe}^{\mrm{u}} ) = 1, \; || \bs{\vDe}^{\mrm{u}} || \leq
r \}$. By Lemma \ref{lemdec},
 \bee
\bb{P}(r) & = & \int_{  0}^r \phi(\ro) \; f_{ || \bs{\vDe}^{\mrm{u}}
|| } (\ro) \; d \ro  =   \int_{  0}^r \phi(\ro) \f{n \ro^{n-1}}{r^n}
d \ro =  \f{n}{r^n} \int_{  0}^r \phi(\ro) \;  \ro^{n-1} \; d \ro.
\eee

\epf

The following two lemmas establish connections between $\phi(.)$, $\bb{P}(.)$ and $\mscr{P}(.)$.

\beL \la{lemm7} For any continuous interval $(a, b)$ with $0 < a < b$,
\[
\int_{  a}^b \phi(\ro) \; d \ro = \f{b \bb{P}(b) - a \bb{P}(a) } { n
} + \f{n-1}{n}  \int_{  a}^b \bb{P}(\ro) \; d \ro.
\]
\eeL

\bpf  By Lemma \ref{Barm}, we have $\bb{P}(r) = \f{n}{r^n} \int_{
0}^r \phi(\ro) \; \ro^{n-1} \; d \ro$.  Since $\phi(\ro)$ is
continuous over $(a, b)$,  we have that $\bb{P}(r)$ is
differentiable with respect to $r$ and that $\phi(\ro) = \f{ \f{
d[\ro^n \bb{P}(\ro)] } { d \ro }  } { n \ro^{n-1} }$ for any $\ro
\in (a, b)$.  Consequently, \bel \int_{  a}^b \phi(\ro) \; d \ro & =
& \int_{ a}^b \f{ \f{ d[\ro^n \bb{P}(\ro)] } { d \ro } } { n
\ro^{n-1} } \;
d \ro \nonumber\\
& = &  \int_{  a}^b \f{ 1 } { n \ro^{n-1} } \; d[\ro^n
\bb{P}(\ro)] \nonumber\\
& = & \lim_{\ep \to 0} \f{ (b - \ep) \bb{P}(b - \ep) - (a + \ep) \bb{P}(a + \ep) } { n } +
\f{n-1}{n}  \int_{  a}^b \bb{P}(\ro) d \ro \la{part}\\
& = & \f{b \bb{P}(b) - a \bb{P}(a) } { n } + \f{n-1}{n}  \int_{
a}^b \bb{P}(\ro) d \ro \la{last} \eel where we have used the
technique of integration by part in (\ref{part}) and the fact that
$\bb{P}(\ro)$ is continuous for any $\ro > 0$ in (\ref{last}).\epf

\beL \la{lemm8}
For any continuous interval $(a, b)$ with $0 < a < b$,
\[
\int_{  0}^r \phi(\ro) \; \ro^{n-1} \; d \ro = [b^n \; \mscr{P}(b) -
a^n \; \mscr{P}(a)]  - (n-1) \int_{  a}^b \mscr{P}(\ro) \; \ro^{n-1}
d \ro.
\]

 \eeL

\bpf By Lemma \ref{chen}, we have $\mscr{P}(\ro) = \f{1}{r} \int_0^r
\phi(\ro) \; d \ro$.  Since $\phi(\ro)$ is continuous over $(a, b)$,
we have that $\mscr{P}(\ro)$ is differentiable with respect to $\ro$
and that $\phi(\ro) = \f{ d[\ro \; \mscr{P}(\ro)] } { d \ro }$ for
any $\ro \in (a, b)$. Hence, \bee  \int_0^r \ro^{n-1} \phi(\ro) \; d
\ro
 & = & \int_{  a}^b  \ro^{n-1} d[\ro \;
\mscr{P}(\ro)]\\
& = & \lim_{\ep \to 0} \li [ (b - \ep)^n \; \mscr{P}(b - \ep) - (a +
\ep)^n \; \mscr{P}(a + \ep) \ri ]- \int_{  a}^b \ro \; \mscr{P}(\ro)
\; (n-1) \ro^{n-2} d
\ro\\
& = & [b^n \; \mscr{P}(b) - a^n \; \mscr{P}(a)]  - (n-1) \int_{
a}^b \mscr{P}(\ro) \; \ro^{n-1} d \ro \eee where we have used the
technique of integration by part and the fact that $\mscr{P}(\ro)$
is continuous for any $\ro > 0$.

\epf

\beL \la{lemm9} Let $q \leq a < b \leq r$.  Then, $|b \bb{P}(b) - a
\bb{P}(a)| \leq \li ( \f{ n r } {q}  + 1 \ri ) (b - a)$.  \eeL

\bpf Note that, for $q \leq a < b \leq r$, we have \bee |b \bb{P}(b) - a \bb{P}(a)| & = & |b \bb{P}(b) - b
\bb{P}(a) +  b \bb{P}(a) - a \bb{P}(a)|\\
& \leq &  b |\bb{P}(b) - \bb{P}(a)| + (b - a) \bb{P}(a) \\
& \leq & b \f{ n(b-a) } { a }  +  (b - a) \leq \li ( \f{ n r } {q} +
1 \ri ) (b - a) \eee where we have used the bound $|\bb{P}(b) -
\bb{P}(a)| \leq \f{ n(b-a) } { a }$, which was derived in the proof
of Theorem $6.1$ in page $856$ of \cite{BLT}. \epf

\beL \la{lemm11} Let $q \leq a < b \leq r$. Then, $| b^n \;
\mscr{P}(b) - a^n \; \mscr{P}(a)| < \li ( \f{2 r^n} { q } +  n
r^{n-1} \ri ) (b-a)$.  \eeL

\bpf

Note that, by Lemma \ref{debound},  $|\mscr{P}(b) - \mscr{P}(a) | \leq \f{2 (b-a)} { a }$, we have \bee | b^n \;
\mscr{P}(b) - a^n \; \mscr{P}(a)| & = &
| b^n \; \mscr{P}(b) - b^n \; \mscr{P}(a) + b^n \; \mscr{P}(a) - a^n \; \mscr{P}(a)| \\
& \leq & b^n |\mscr{P}(b) - \mscr{P}(a) | +  (b^n - a^n)
\mscr{P}(a)\\
& \leq & \f{2 b^n (b-a)} { a }  + (b^n - a^n)\\
& < & \f{2 b^n (b-a)} { a }  + n b^{n-1} (b-a)\\
& = & \li ( \f{2 b^n} { a } +  n b^{n-1} \ri ) (b-a)\\
& \leq & \li ( \f{2 r^n} { q } +  n r^{n-1} \ri ) (b-a) \eee where
we have used the inequality $b^n - a^n < n b^{n-1} (b-a)$ which can
be shown by using Taylor's expansion formula $b^n = a^n + n
\xi^{n-1} (b -a) < a^n + n b^{n-1} (b-a)$ with some $\xi \in (a,
b)$.

\epf

We are now in the position to prove the transform formulas for each cases.

\bed

\item [Case (1)] Let $0 = p_0 < p_1 <
\cd < p_k < p_{k+1} = r$ where $p_1, \cd, p_{k}$ are $k \geq 0$
discontinuities.  By Lemma \ref{lemm7}, we have \bee \int_{0}^r
\phi(\ro) d \ro & = & \lim_{\ep \downarrow 0} \int_{\ep}^r
\phi(\ro) d \ro\\
& = & \lim_{\ep \downarrow 0} \int_{  \ep}^{p_{1}} \phi(\ro) \; d \ro + \sum_{i=1}^k \int_{  p_i}^{p_{i+1}} \phi(\ro) \; d \ro\\
& = & \lim_{\ep \downarrow 0} \li [
\f{p_1 \bb{P}(p_1) - \ep \bb{P}(\ep) } { n } + \f{n-1}{n}  \int_{  \ep}^{p_1} \bb{P}(\ro) d \ro \ri ]\\
&  & + \sum_{i=1}^k \li [ \f{p_{i+1} \bb{P}(p_{i+1}) - p_i
\bb{P}(p_i) } { n } + \f{n-1}{n}  \int_{
p_i}^{p_{i+1}} \bb{P}(\ro) d \ro \ri ]\\
& = & \lim_{\ep \downarrow 0} \li [ \f{ - \ep \bb{P}(\ep) } { n } +
\f{n-1}{n}  \int_{  \ep}^{p_1} \bb{P}(\ro) d \ro \ri ] + \f{r
\bb{P}(r) } { n } + \f{n-1}{n}  \int_{p_1}^r \bb{P}(\ro) d \ro. \eee
Since $0 \leq \bb{P}(\ro) \leq 1, \; \fa \ro > 0$, we have
$\lim_{\ep \downarrow 0} \ep \bb{P}(\ep) = 0$ and $\lim_{\ep
\downarrow 0} \int_{ \ep}^{p_1} \bb{P}(\ro) d \ro = \int_{  0}^{p_1}
\bb{P}(\ro) d \ro$. It follows that $\int_{0}^r \phi(\ro) d \ro =
\f{r \bb{P}(r) } { n } + \f{n-1}{n}  \int_{0}^r \bb{P}(\ro) d \ro$
and that $\mscr{P}(r) = \f{1}{r} \int_{0}^r \phi(\ro) d \ro =
\f{\bb{P}(r) } { n } + \f{n-1}{n r}  \int_{0}^r \bb{P}(\ro) d \ro$.

By Lemma \ref{lemm8} and similar techniques, we can show the
expression for $\bb{P}(r)$ in this case.

\item [Case (2)] In this case, the discontinuities can be represented as a monotone decreasing sequence $\{p_i\}_{i=1}^\iy$ such
that $r = p_0 > p_1 > p_2 > \cd > p_k > \cd$ and $\lim_{k \to \iy} p_k = 0$.  By Lemma \ref{lemm7}, we have \bee
\int_{0}^r \phi(\ro) d \ro
& = & \lim_{k \to \iy} \sum_{i=1}^k \int_{  p_i}^{p_{i-1}} \phi(\ro) \; d \ro\\
& = & \lim_{k \to \iy} \sum_{i=1}^k \li [ \f{p_{i-1} \bb{P}(p_{i-1}) - p_i \bb{P}(p_i) } { n } + \f{n-1}{n}
\int_{  p_i}^{p_{i-1}} \bb{P}(\ro) d \ro \ri ]\\
& = & \lim_{k \to \iy} \li [ \f{r \bb{P}(r) - p_k \bb{P}(p_k)}{n}  +
\f{n-1}{n} \int_{  p_k}^{r} \bb{P}(\ro) d \ro \ri ]. \eee Since
{\small $0 \leq \bb{P}(\ro) \leq 1, \; \fa \ro > 0$} and {\small
$\lim_{k \to \iy} p_k = 0$}, we have {\small $\lim_{k \to \iy} p_k
\bb{P}(p_k) = 0$} and {\small $\lim_{k \to \iy} \int_{p_k}^{r}
\bb{P}(\ro) d \ro  = \int_{0}^{r} \bb{P}(\ro) d \ro$}.  It follows
that $\int_{0}^r \phi(\ro) d \ro = \f{r \bb{P}(r) } { n } +
\f{n-1}{n} \int_{0}^r \bb{P}(\ro) d \ro$ and $\mscr{P}(r) = \f{1}{r}
\int_{0}^r \phi(\ro) d \ro = \f{\bb{P}(r) } { n } + \f{n-1}{n r}
\int_{0}^r \bb{P}(\ro) d \ro$.

By Lemma \ref{lemm8} and similar techniques, we can show the expression for $\bb{P}(r)$ in this case.

\item [Case (3)] In this case, let $r_*$ be the smallest positive cluster point.
Let $q = \f{r_*  } {2}$. We can write $\int_{0}^r \phi(\ro) d \ro =
\int_{0}^{q} \phi(\ro) d \ro + \int_{q}^r \phi(\ro) d \ro$. Applying
the result of Case (2), we have $\int_{0}^{q} \phi(\ro) d \ro = \f{q
\bb{P}(q) } { n } + \f{n-1}{n} \int_{0}^q \bb{P}(\ro) d \ro$.  We
consider $\int_{q}^r \phi(\ro) d \ro$.  For any $\vep > 0$, by Lemma
\ref{lemm4}, we can write \be \la{eq10} \int_{q}^r \phi(\ro) d \ro =
\sum_{(a, b) \in \wh{\mscr{I}_\vep} }  \int_{(a, b)} \phi(\ro) d \ro
+  \sum_{(a, b) \in \mscr{I}_\vep } \int_{(a, b)}  \phi(\ro) d \ro
\ee where $\int_{(a, b)}$ means the integration over interval $(a,
b)$ and $\sum_{(a, b) \in \wh{\mscr{I}_\vep} }$ means the summation
over all intervals of $\wh{\mscr{I}_\vep}$.  The notion of
$\sum_{(a, b) \in \mscr{I}_\vep }$ is similar.

 To evaluate $\sum_{(a, b) \in
\wh{\mscr{I}_\vep} } \int_{(a, b)} \phi(\ro) d \ro$, we arrange the
intervals in $\wh{\mscr{I}_\vep}$ as $(a_i, b_i), \; i = 1, \cd, k$
such that $a_1 = q, \; b_i < a_{i+1}, \; i = 1, \cd, k - 1$ (Here
$k$ is the total number of intervals). Note that, by Lemma
\ref{lemm7}, \bel \sum_{(a, b) \in \wh{\mscr{I}_\vep} } \int_{(a,
b)} \phi(\ro) d \ro & = & \sum_{i = 1}^k \li [  \f{b_i \bb{P}(b_i) -
a_i \bb{P}(a_i) } { n } + \f{n-1}{n}
\int_{  a_i}^{b_i} \bb{P}(\ro) d \ro \ri ] \nonumber\\
& = &  \f{r \bb{P}(r) - q \bb{P}(q) } { n } + \f{n-1}{n} \int_{
q}^{r} \bb{P}(\ro) d \ro \la{eq11}\\
&   & - \sum_{i = 1}^{k-1} \li [ \f{a_{i+1} \bb{P}(a_{i+1}) - b_i
\bb{P}(b_i) } { n } + \f{n-1}{n} \int_{ b_i}^{a_{i+1}} \bb{P}(\ro) d
\ro \ri ] \nonumber. \eel By Lemma \ref{lemm9}, we have $|a_{i+1}
\bb{P}(a_{i+1}) - b_i \bb{P}(b_i)| < \li ( \f{ n r } {q}  + 1 \ri )
( a_{i+1} - b_i), \;\; i = 1, \cd, k-1$ and \bel \sum_{i = 1}^{k-1}
\li |  \f{a_{i+1} \bb{P}(a_{i+1}) - b_i \bb{P}(b_i) } { n } \ri | &
< & \sum_{i =
1}^{k-1} \li [ \li ( \f{ n r } {q}  + 1 \ri ) ( a_{i+1} - b_i) \ri ]  =  \li ( \f{ n r } {q}  + 1 \ri ) \sum_{i = 1}^{k-1}  ( a_{i+1} - b_i) \nonumber\\
& = & \li ( \f{ n r } {q}  + 1 \ri ) \vep \la{eq12}. \eel Since $0
\leq \bb{P}(\ro) \leq 1$, we have \be \la{eq13} \sum_{i = 1}^{k-1}
\li | \f{n-1}{n} \int_{  b_i}^{a_{i+1}} \bb{P}(\ro) d \ro \ri | \leq
\f{n-1}{n} \sum_{i = 1}^{k-1} ( a_{i+1} - b_i) = \f{n-1}{n} \vep.
\ee By (\ref{eq11}), (\ref{eq12}), and (\ref{eq13}), \bel  \li |
\sum_{i = 1}^{k-1} \li [ \f{a_{i+1} \bb{P}(a_{i+1}) - b_i
\bb{P}(b_i) } { n } + \f{n-1}{n} \int_{  b_i}^{a_{i+1}} \bb{P}(\ro)
d \ro \ri ] \ri |
& < &  \li ( \f{ n r } {q}  + 1 \ri ) \vep + \f{n-1}{n} \vep \nonumber\\
& = & \li ( \f{ n r } {q}  + 1 + \f{n-1}{n} \ri ) \vep . \la{eq14} \eel Now we bound $\sum_{(a, b) \in
\mscr{I}_\vep } \int_{(a, b)} \phi(\ro) d \ro$.  By Lemmas \ref{lemm7} and \ref{lemm9}, \bel \sum_{(a, b) \in
\mscr{I}_\vep } \int_{(a, b)} \phi(\ro) d \ro & =  & \sum_{(a, b) \in \mscr{I}_\vep } \li [ \f{b \bb{P}(b) - a
\bb{P}(a) } { n
} + \f{n-1}{n}  \int_{  a}^b \bb{P}(\ro) d \ro \ri ] \nonumber\\
& < &  \sum_{(a, b) \in \mscr{I}_\vep } \li [  \li ( \f{ n r } {q}  + 1 \ri ) (b - a) + \f{n-1}{n}  (b - a) \ri
] \nonumber\\
& = &  \li ( \f{ n r } {q}  + 1 + \f{n-1}{n} \ri ) \sum_{(a, b) \in \mscr{I}_\vep } (b - a) \nonumber\\
& = & \li ( \f{ n r } {q}  + 1 + \f{n-1}{n} \ri ) \vep. \la{eq15}
\eel Therefore, by (\ref{eq10}), (\ref{eq11}), (\ref{eq14}) and
(\ref{eq15}), \bee &  & \li |  \int_{q}^r \phi(\ro) d \ro  - \li [
\f{r \bb{P}(r) - q \bb{P}(q) } { n } + \f{n-1}{n} \int_{  q}^{r}
\bb{P}(\ro) d \ro \ri ] \ri
|\\
& \leq & \li | \sum_{i = 1}^{k-1} \li [ \f{a_{i+1} \bb{P}(a_{i+1}) - b_i \bb{P}(b_i) } { n } + \f{n-1}{n}
\int_{  b_i}^{a_{i+1}} \bb{P}(\ro) d \ro \ri ] \ri |  + \sum_{(a, b) \in \mscr{I}_\vep } \int_{(a, b)} \phi(\ro) d \ro\\
& < & 2  \li ( \f{ n r } {q}  + 1 + \f{n-1}{n} \ri ) \vep. \eee
Since the above argument holds for arbitrarily small $\vep > 0$, it
must be true that $\int_{q}^r \phi(\ro) \; d \ro = \f{r \bb{P}(r) -
q \bb{P}(q) } { n } + \f{n-1}{n} \int_{  q}^{r} \bb{P}(\ro) d \ro$.
It follows that \bee \int_{0}^r \phi(\ro) d \ro & = & \int_0^{q} \phi(\ro) d \ro + \int_{q}^r \phi(\ro) d \ro\\
& = &  \f{q \bb{P}(q) } { n } + \f{n-1}{n} \int_{0}^q \bb{P}(\ro) d \ro + \f{r \bb{P}(r) - q \bb{P}(q) } { n } + \f{n-1}{n} \int_{  q}^{r} \bb{P}(\ro) d \ro\\
& = & \f{r \bb{P}(r) } { n } + \f{n-1}{n} \int_{  0}^{r} \bb{P}(\ro)
d \ro,\eee leading to the formula for $\mscr{P}(r)$.

To show the formula for $\bb{P}(r)$, recall that $r^n \bb{P}(r) = n
\int_{  0}^r \phi(\ro) \; \ro^{n-1} \; d \ro$.  We write \be \int_{
0}^r \phi(\ro)  \ro^{n-1} \; d \ro = \int_{  0}^q \phi(\ro)
\ro^{n-1} \; d \ro + \int_{  q}^r \phi(\ro) \ro^{n-1} \; d \ro.
\la{sep} \ee By Lemma \ref{lemm4}, we can write \be \int_{ q}^r
\phi(\ro)  \ro^{n-1} \; d \ro  = \sum_{(a,b) \in \wh{\mscr{I}_\vep}
} \int_{(a, b)} \phi(\ro) \ro^{n-1} \; d \ro + \sum_{(a,b) \in
\mscr{I}_\vep } \int_{(a, b)} \phi(\ro) \ro^{n-1} \; d \ro.
\la{sep2} \ee To evaluate $\sum_{(a, b) \in \wh{\mscr{I}_\vep} }
\int_{(a, b)} \phi(\ro) \ro^{n-1} d \ro$, we arrange the intervals
in $\wh{\mscr{I}_\vep}$ as $(a_i, b_i), \; i = 1, \cd, k$ such that
$a_1 = q, \; b_i < a_{i+1}, \; i = 1, \cd, k - 1$ (Here $k$ is the
total number of intervals). Note that, by Lemma \ref{lemm8},
 \bel  \sum_{(a, b) \in \wh{\mscr{I}_\vep}
} \int_{(a, b)} \phi(\ro) \ro^{n-1} d \ro
 & = & \sum_{i = 1}^k \li [  b_i^n \mscr{P}(b_i) - a_i^n
\mscr{P}(a_i)
- (n-1) \int_{  a_i}^{b_i} \mscr{P}(\ro) \ro^{n-1} d \ro \ri ] \nonumber\\
& = &  r^n \mscr{P}(r) - q^n \mscr{P}(q)  - (n-1) \int_{  q}^{r} \mscr{P}(\ro) \ro^{n-1} d \ro \la{eq17}\\
&   & - \sum_{i = 1}^{k-1} \li [ a_{i+1}^n \mscr{P}(a_{i+1}) - b_i^n
\mscr{P}(b_i) - (n-1) \int_{ b_i}^{a_{i+1}} \mscr{P}(\ro) \ro^{n-1}
d \ro \ri ]. \nonumber\eel By Lemma \ref{lemm11}, we have $|
a_{i+1}^n \; \mscr{P}(a_{i+1}) - b_i^n \; \mscr{P}(b_i)| < \li (
\f{2 r^n} { q } +  n r^{n-1} \ri ) (a_{i+1} - b_i)$.  Hence, \bel
\li | \sum_{i = 1}^{k-1} \li [ a_{i+1}^n \mscr{P}(a_{i+1}) - b_i^n
\mscr{P}(b_i) \ri ] \ri | & < &
\li ( \f{2 r^n} { q } +  n r^{n-1} \ri ) \sum_{i = 1}^{k-1} (a_{i+1} - b_i) \nonumber\\
& = &  \li ( \f{2 r^n} { q } +  n r^{n-1} \ri ) \vep. \la{eq18} \eel
On the other hand, observing that $\int_{  a}^b \mscr{P}(\ro) \;
\ro^{n-1} d \ro < r^{n-1} (b-a)$, we have \be \sum_{i = 1}^{k-1}
\int_{  b_i}^{a_{i+1}} \mscr{P}(\ro) \ro^{n-1} d \ro < r^{n-1}
\sum_{i = 1}^{k-1} (a_{i+1} - b_i) = r^{n-1} \vep. \la{eq19} \ee By
(\ref{eq17}), (\ref{eq18}) and (\ref{eq19}), \bel &  & \li |
\sum_{(a, b) \in \wh{\mscr{I}_\vep} } \int_{(a, b)} \phi(\ro)
\ro^{n-1} d \ro - \li [ r^n
\mscr{P}(r) - q^n \mscr{P}(q)  - (n-1) \int_{  q}^{r} \mscr{P}(\ro) \ro^{n-1} d \ro \ri ] \ri | \nonumber\\
& < &  \li ( \f{2 r^n} { q } +  n r^{n-1} \ri ) \vep - (n-1) r^{n-1} \vep \nonumber\\
& = & \li ( \f{2 r^n} { q } +  r^{n-1} \ri ) \vep. \nonumber \\
\la{eqB12} \eel

Now we bound $\sum_{(a, b) \in \mscr{I}_\vep } \int_{(a, b)} \phi(\ro) \ro^{n-1} d \ro$.  By Lemmas \ref{lemm8}
and \ref{lemm11}, \bel \sum_{(a, b) \in \mscr{I}_\vep } \int_{(a, b)} \phi(\ro) \ro^{n-1} d \ro & =  & \sum_{(a,
b) \in \mscr{I}_\vep } \li [ b^n \mscr{P}(b) -
a^n \mscr{P}(a) - (n-1) \int_{  a}^b \mscr{P}(\ro) \ro^{n-1} d \ro \ri ] \nonumber\\
& < &  \sum_{(a, b) \in \mscr{I}_\vep } \li [  \li ( \f{ 2 r^n } {q}  + n r^{n-1} \ri ) (b - a) - (n-1) r^{n-1}
(b - a) \ri
] \nonumber\\
& = &  \li ( \f{ 2 r^n } {q}  +  r^{n-1} \ri ) \sum_{(a, b) \in \mscr{I}_\vep } (b - a) \nonumber\\
& = & \li ( \f{ 2 r^n } {q}  +  r^{n-1} \ri ) \vep. \la{eqB13}\eel
Therefore, by (\ref{sep2}), (\ref{eqB12}) and (\ref{eqB13}),
 \bee & & \li | \int_q^r \phi(\ro) \ro^{n-1} d \ro - \li [ r^n
\mscr{P}(r) - q^n \mscr{P}(q)  - (n-1) \int_{  q}^{r} \mscr{P}(\ro) \ro^{n-1} d \ro \ri ] \ri |\\
& < & \li ( \f{2 r^n} { q } +  r^{n-1} \ri ) \vep + \li ( \f{ 2 r^n
} {q}  +  r^{n-1} \ri ) \vep =  2 \li ( \f{ 2 r^n } {q}  + r^{n-1}
\ri ) \vep. \eee Since the argument applies to arbitrarily small
$\vep
> 0$, it must be true that $\int_q^r \phi(\ro) \ro^{n-1} d \ro = \li [ r^n \mscr{P}(r) - q^n
\mscr{P}(q)  - (n-1) \int_{  q}^{r} \mscr{P}(\ro) \ro^{n-1} d \ro
\ri ]$.  Therefore, \bee \int_0^r \phi(\ro) \ro^{n-1} d \ro
& = & \int_0^q \phi(\ro) \ro^{n-1} d \ro + \int_q^r \phi(\ro) \ro^{n-1} d \ro\\
& = & q^n \mscr{P}(q)  - (n-1) \int_{  0}^{q} \mscr{P}(\ro) \ro^{n-1} d \ro +  r^n \mscr{P}(r) - q^n \mscr{P}(q)  - (n-1) \int_{  q}^{r} \mscr{P}(\ro) \ro^{n-1} d \ro\\
& = & r^n \mscr{P}(r)  - (n-1) \int_{  0}^{r} \mscr{P}(\ro)
\ro^{n-1} d \ro, \eee from which we find the formula for
$\bb{P}(r)$.

\item [Case (4)] In this case, let $r_*$ be the smallest positive cluster point.
Let $q = \f{r_*  } {2}$. We can write $\int_{0}^r \phi(\ro) d \ro =
\int_{0}^{q} \phi(\ro) d \ro + \int_{q}^r \phi(\ro) d \ro$. Applying
the result of Case (1), we have $\int_{0}^{q} \phi(\ro) d \ro = \f{q
\bb{P}(q) } { n } + \f{n-1}{n} \int_{0}^q \bb{P}(\ro) d \ro$.  By a
method similar to that of Case (3), we have $\int_{q}^r \phi(\ro) \;
d \ro = \f{r \bb{P}(r) - q \bb{P}(q) } { n } + \f{n-1}{n} \int_{
q}^{r} \bb{P}(\ro) \; d \ro$.  Combining the two integrals gives the
formula for $\mscr{P}(r)$.  The proof for the formula of $\bb{P}(r)$
is similar.

\eed

\section{Proofs of Theorem 3 and 4}

For completeness of argument, we need to quote a general complexity result established in \cite{Chen_SIAM} as
Theorem \ref{Bound_General} at below. This theorem concerns the sampling complexity of the Sample Reuse
Algorithm proposed in page 1963 of \cite{C0}.

 \beT \la{Bound_General} Let $d$ be the dimension of uncertainty parameter
space. Then, for arbitrary gridding scheme, the equivalent number of
grid points based on the Sample Reuse Algorithm \cite{C0} is
strictly bounded from above by $1 + d \; \ln \lm$, i.e.,
$m_{\mrm{eq}} < 1 + d \; \ln \lm$. \eeT

\bpf We first establish the following inequality (\ref{inc}) that will be used to prove Theorem
\ref{Bound_General}.

 \be \la{inc} \frac{1}{x} + \ln x > 1, \qqu \fa x > 1. \ee

To prove \ref{inc}, let $f(x) = \frac{1}{x} + \ln x$.  Then $f(1) =
1$ and $\frac{d \; f(x) } {d x } = \frac{x - 1}{x^2} > 0, \; \forall
x
> 1$.  It follows that $f(x) > 1, \; \forall x > 1$.

Now we are in the position to prove Theorem \ref{Bound_General}.
Observing that $\li ( \frac{r_m}{r_1} \ri )^d = \prod_{i=1}^{m-1}
\li ( \frac{r_{i+1}}{r_i} \ri )^d$, we have $\ln \left(
\frac{r_m}{r_1} \right )^d = \sum_{i=1}^{m-1} \ln \left (
\frac{r_{i+1}}{r_i} \right )^d$.  Therefore, \begin{eqnarray*}
\sum_{i=1}^{m-1} \li( \frac{r_i}{r_{i+1}} \ri )^d + \ln \left(
\frac{r_m}{r_1} \right )^d
 & = & \sum_{i=1}^{m-1} \left [ \frac{1 } { \li ( \frac{r_{i+1}}{r_i} \ri )^d} + \ln \left (
\frac{r_{i+1}}{r_i} \right )^d \right ].
\end{eqnarray*} Since $\li( \frac{r_{i+1}}{r_i} \ri )^d > 1, \; i = 1,
\cdots, m-1$, it follows from (\ref{inc}) that $\frac{1 } { \li (
\frac{r_{i+1}}{r_i} \ri )^d } + \ln \left ( \frac{r_{i+1}}{r_i}
\right )^d > 1$ for $i = 1, \cdots , m-1$.  Hence, $\sum_{i=1}^{m-1}
\li ( \frac{r_i}{r_{i+1}} \ri )^d + \ln \left( \frac{r_m}{r_1}
\right )^d
>m - 1$,
or equivalently, $m - \sum_{i=1}^{m-1} \li ( \frac{r_i}{r_{i+1}} \ri
)^d < 1 + \ln \left( \frac{r_m}{r_1} \right )^d = 1 + d \ln \lm$.

Finally, by Theorem 1 of \cite{C0} and the definition of
$m_{\mrm{eq}}$, we have $m_{\mrm{eq}} = m - \sum_{i=1}^{m-1} \li (
\frac{r_i}{r_{i+1}} \ri )^d < 1 + d \ln \lm$.

\epf

\subsection{Proof of Theorem \ref{grid_uni}}

By Lemma \ref{chen}, $|\mscr{P} ( r ) - \mscr{P}^* ( r)| \leq \f{ 2
\; (r_{i+1} - r_i) } {r_i}, \qu \fa r \in [r_i, r_{i+1}]$. Thus, it
suffices to show $\f{ 2 \; (r_{i+1} - r_i) } { r_i} < \ep$, i.e.,
\be \la{con3} \f{ r_{i+1} }{ r_i } < 1 + \f{\ep}{2}. \ee By the
definition of uniform griding, for $i = 1, \cd, m-1$, \bee \f{
r_{i+1} } {r_i} & = & \f{ a - \f{ (m - i - 1) (\lm - 1) } { (m -1)
\lm } a } { a - \f{ (m - i) (\lm - 1) }{ (m -1) \lm } a } =  1 + \f{
\lm - 1 } { m- 1 + (\lm -1) (i -1) } \leq  1 + \f{ \lm - 1 } { m - 1
}. \eee By virtue of (\ref{con3}), to guarantee that the gridding
error is less than $\ep$, it suffices to ensure $1 + \f{ \lm - 1 } {
m - 1 } < 1 + \f{\ep}{2}$, i.e., $m > 1 + \f{2 (\lm -1)}{ \ep}$.
Hence, it suffices to have $m \geq 2 + \li \lf \f{ 2(\lm -1) } { \ep
} \ri \rf$.  It can be verified that $\f{ r_i } { r_{i+1} } = 1 -
\f{ 1 } { \f{m-1} {\lm -1} + i }$ for $i = 1, \cd, m-1$.

Let $\bs{n}^k$ be the total number of simulations on the direction
associated with directional sample $U^k, \; k = 1, \cd, N$. Applying
Theorem 1 of \cite{C0} and  Theorem \ref{Bound_General} in this
paper to a sample reuse process conditioned upon a direction with
grid points $r_1, \cd, r_m$ and sample size $N = 1$, we have $\bb{E}
[\bs{n}^k \mid U^k] = m - \sum_{i=1}^{m-1} \frac{r_i}{r_{i+1}} < 1 +
d \ln \lm$  and consequently $\bb{E} [\bs{n}^k] = \bb{E}[ \bb{E}
[\bs{n}^k \mid U^k] ] = m - \sum_{i=1}^{m-1} \frac{r_i}{r_{i+1}} < 1
+ d \ln \lm$ for $k = 1, \cd , N$. Finally, the proof is completed
by invoking the definition of equivalent number of grid points.

\subsection{Proof of Theorem \ref{Grid_geometric}}

By the definition of uniform griding, we have $\f{ r_{i+1} }{ r_i }
= \lm^{ \f{1}{m-1} }$. Hence, by (\ref{con3}), it suffices to show
$\lm^{ \f{1}{m-1} } < 1 + \f{\ep}{2}$, which can be reduced to $m >
1 + \f{ \ln \lm } { \ln \li ( 1 + \f{\ep}{2} \ri ) }$. This
inequality is equivalent to $m \geq 2 + \li \lf \f{ \ln \lm } { \ln
\li ( 1 + \f{\ep}{2} \ri ) } \ri \rf$.  By letting $\bs{n}^k$ be the
total number of simulations on the direction associated with
directional sample $U^k, \; k = 1, \cd, N$ and applying Theorem 1 of
\cite{C0} and Theorem \ref{Bound_General} in this paper to a sample
reuse process conditioned upon a direction with grid points $r_1,
\cd, r_m$ and sample size $N = 1$, we have $\bb{E} [\bs{n}^k \mid
U^k] = m - (m-1) \left( \frac{1}{\lm} \right)^{ \f{1}{m -1} } < 1 +
d \ln \lm$ and consequently $\bb{E} [\bs{n}^k] = \bb{E}[ \bb{E}
[\bs{n}^k \mid U^k] ] = m - (m-1) \left( \frac{1}{\lm} \right)^{
\f{1}{m -1} } < 1 + d \ln \lm$ for $k = 1, \cd , N$. The proof is
completed by using the definition of equivalent number of grid
points.


\begin{thebibliography}{10}

\bibitem{bai}E. W. BAI, R. TEMPO, AND M. FU,
``Worst-case properties of the uniform distribution and randomized
algorithms for robustness analysis,'' {\it Mathematics of Control,
Signals and Systems}, vol. 11, pp. 183-196, 1998.

\bibitem{BL}
B. R. BARMISH AND C. M. LAGOA, ``The uniform distribution: a
rigorous justification for its use in robustness analysis,'' {\it
Mathematics of Control, Signals and Systems},
 vol. 10 (1997), pp. 203-222.


\bibitem{BLT} B. R. BARMISH, C. M. LAGOA, AND R. TEMPO,
``Radially truncated uniform distributions for probabilistic robustness of control systems,'' {\it Proc. of
American Control Conference}, pp. 853-857, Albuquerque, New Mexico, June 1997.


\bibitem{Barmish2} B. R. BARMISH AND P. S. SHCHERBAKOV, ``On avoiding
vertexization of robustness problems: The approximate feasibility concept,'' {\it IEEE Transactions on Automatic
Control}, vol. 42, pp. 819-824, 2002.

\bibitem{C0} X. CHEN, K. ZHOU, AND J. ARAVENA, ``Fast construction of
robustness degradation function,'' {\it SIAM Journal on Control and Optimization}, vol. 42, pp. 1960-1971, 2004.

\bibitem{C2} X. CHEN, K. ZHOU, AND J. ARAVENA, ``Fast universal algorithms for robustness analysis,''
{\it Proceedings of IEEE Conference on Decision and Control}, pp. 1926-1931, Maui, Hawaii, December 2003.

\bib{Chen_SIAM} X. CHEN, K. ZHOU, AND J. ARAVENA, ``Probabilistic robustness analysis --- Risks, complexity and
algorithms,'' submitted for publication.

\bib{Chernoff} H. CHERNOFF, ``A measure of asymptotic
efficiency for tests of a hypothesis based on the sum of observations,'' {\it Annals of Mathematical
Statistics}, vol. 23, pp. 493-507, 1952.

\bib{Doyle} J. C. DOYLE AND G. STEIN, ``Multivariable feedback
design: concepts for a classical/modern synthesis,'' {\em IEEE
Trans. Autom. Control}, vol. 26, pp. 4-16, 1981.

\bibitem{GS} R. R DE GASTON AND M. G. SAFONOV,
``Exact calculation of the multiloop stability margin,'' {\em IEEE
Trans. Autom. Control}, vol. 33, pp. 156-171, 1988.



\bib{Kan} S. KANEV, B. De SCHUTTER, AND M. VERHAEGEN, ``An ellipsoid
algorithm for probabilistic robust controller design,'' {\it Systems and Control Letters}, vol. 49, pp. 365-375,
2003.



\bib{Ko} V. KOLTCHINSKII, C.T. ABDALLAH, M. ARIOLA, P. DORATO, AND D.
PANCHENKO, ``Improved sample complexity estimates for statistical learning control of uncertain systems,'' {\it
IEEE Transactions on Automatic Control}, vol. 46, pp. 2383-2388, 2000.

\bib{Qiu} L. QIU, B. BERNHARDSSON, A. RANTZER, E. J. DAVISON, P.
M. YOUNG, AND J. C. DOYLE, ``A formula for computation of the real stability radius,'' vol. 31, pp. 879-890,
1995.

\bibitem{SR}
R. F. STENGEL AND L. R. RAY, ``Stochastic robustness of linear time-invariant systems,'' {\it IEEE Transaction
on Automatic Control}, vol. 36, pp. 82-87, 1991.


\bib{Wang} Q. WANG AND R. F. STENGEL, ``Robust control of nonlinear
systems with parametric uncertainty,'' {\it Automatica}, vol. 38, pp. 1591-1599, 2002.

\bibitem{ZDG} K. ZHOU, J. C. DOYLE, AND K. GLOVER,
{\it Robust and Optimal Control}, Prentice Hall, Upper Saddle River, NJ, 1996.

\end{thebibliography}
\end{document}